\documentclass[11pt,tightenlines,superscriptaddress,floatfixolumn,english,aps,notitlepage,nofootinbib]{revtex4-1}
\usepackage{amsmath}
\usepackage{esint}   
\usepackage{graphicx}
\usepackage{hyperref}
\usepackage{color}
\usepackage{xcolor}
\usepackage{enumerate}
\usepackage[utf8]{inputenc} 
\advance\voffset -0.2in

\newcommand{\be}{\begin{eqnarray}}
\newcommand{\ee}{\end{eqnarray}}
\newcommand{\non}{\nonumber\\}

\newcommand{\ave}[1]{\left\langle #1 \right\rangle}

\newcommand{\mev}{{\rm \, MeV}}
\newcommand{\gev}{{\rm \, GeV}}

\newcommand{\mh}{\hat{\mu}_B}
\newcommand{\mb}{\bar{\mu}_B}
\newcommand{\dN}{{\cal N}}
\newcommand{\dM}{{\cal M}}
\newcommand{\cum}{K}
\newcommand{\neve}{n_{\rm events}}

\definecolor{vkcolor2}{HTML}{336600}

\begin{document}

\title{Net-baryon multiplicity distribution consistent with lattice QCD}

\author{Adam Bzdak}
\email{bzdak@fis.agh.edu.pl}
\affiliation{AGH University of Science and Technology,
Faculty of Physics and Applied Computer Science,
30-059 Krak\'ow, Poland}

\author{Volker Koch}
\email{vkoch@lbl.gov}
\affiliation{Nuclear Science Division, 
Lawrence Berkeley National Laboratory, 
Berkeley, CA, 94720, USA}

\begin{abstract}
We determine the net-baryon
multiplicity distribution which reproduces all cumulants measured so far by lattice QCD. We
present the dependence on the volume and temperature of this distribution. We find that for 
temperatures and volumes encountered in heavy ion reactions, the multiplicity distribution is very
close to the Skellam distribution, making the experimental determination of it rather
challenging. We further provide estimates for the statistics required to measure cumulants of the
net-baryon and net-proton distributions.   
\end{abstract}

\maketitle

\section{Introduction}

In recent years is has been realized that the fluctuations of conserved charges provide experimental
access to the phase structure of QCD
\cite{Jeon:2000wg,Asakawa:2000wh,Stephanov:2008qz,Skokov:2010uh,Stephanov:2011pb,Luo:2011rg,
Luo:2017faz,Herold:2016uvv,Zhou:2012ay,Wang:2012jr,Karsch:2011gg,Schaefer:2011ex,Chen:2011am,
Fu:2009wy,Cheng:2008zh}. In particular the cumulants of the net-baryon number distribution
have been found to be sensitive to the details of the transition from hadron gas to the quark-gluon 
plasma. Not only are they  sensitive probes for  a possible critical point \cite{Stephanov:2008qz}
but they may also reveal the chiral critical behavior underlying  the cross-over transition
\cite{Aoki:2006we,Borsanyi:2010cj,Bazavov:2011nk} at vanishing net-baryon chemical potential
\cite{Skokov:2010uh,Karsch:2011gg}. This insight has motivated the experimental measurement of these
cumulants \cite{Adamczyk:2013dal,Rustamov:2017lio}. Since the detection of neutrons is difficult if not
impossible in these experiments, the present  measurements are restricted to net-proton
cumulants, which however, can be related to the baryon cumulants  \cite{Kitazawa:2011wh,Kitazawa:2012at}.
Also, the experimental data need to be corrected for global baryon number conservation
\cite{Bzdak:2012an} and detector specific effects such as efficiencies
etc. \cite{Bzdak:2012ab,Luo:2014rea,Bzdak:2016qdc,Braun-Munzinger:2015hba,Braun-Munzinger:2016yjz,Kitazawa:2016awu,He:2018mri,Nonaka:2018mgw}.

The cumulants of the net-baryon number distribution for a system at vanishing net-baryon chemical
potential can be calculated from lattice QCD and at present results are available up the eighth
order, $\cum_{8}$ \cite{Bazavov:2017dus,Borsanyi:2018grb}. However, so far it has not been
possible to extract the full net-baryon multiplicity distribution directly from lattice QCD. The
knowledge of this multiplicity distribution would be not only of scientific interest but also has
some practical importance. 
For example, one can calculate to which extent the multiplicity distribution consistent with lattice
QCD deviates from that of uncorrelated
particles, the Skellam distribution. 
In addition, one can
determine how far one has to sum over the multiplicity distribution in order to get a
reasonable approximation for a given cumulant. Furthermore, as we shall discuss in
some detail in the Appendix, one can apply the delta method
\cite{davison2003statistical,Luo:2014rea,Luo:2017faz} in order to estimate the needed
statistics for the measurement of net-baryon cumulants.

The net-baryon number multiplicity distribution has been determined based on an effective model of QCD \cite{Morita:2013tu} as well as from the hadron resonance gas model \cite{BraunMunzinger:2011dn}.
In this paper we will construct the net-baryon multiplicity distribution based on lattice QCD
results \cite{Vovchenko:2017xad} and on the cluster expansion model (CEM) described in
\cite{Vovchenko:2017gkg}. As demonstrated in \cite{Vovchenko:2017gkg,Vovchenko:2018zgt} the CEM reproduces all
net-baryon number cumulants determined by lattice QCD so far. Consequently the cumulants of the
constructed multiplicity distribution agree with all cumulants presently known from lattice QCD (up
to eighth order). Thus, we consider the extracted multiplicity
distribution as the best presently available template. We note that an alternative virial expansion
has been proposed in \cite{Almasi:2018lok}, which also fits the presently available lattice
data within errors. While both expansions have quite different asymptotic behavior of the virial coefficients,
the resulting multiplicity distributions are very similar as we shall discuss. 

This paper is organized as follows. In the next section be briefly review the CEM and show how it
can be used to extract the net-baryon multiplicity distribution. Next, we present the results and
finish with a discussion, which includes an estimate of the required statistics for the measurement
of net-baryon and net-proton cumulants.

\section{Net-baryon multiplicity distribution}

To illustrate how one can extract the multiplicity distribution, let us start 
with the fugacity expansion of the pressure
\begin{align}
  \frac{P}{T^{4}}=\frac{1}{VT^{3}}\ln(Z)=\sum_{k=0}^{\infty}p_{k}(T)\cosh\left( k \mh \right) ,
  \label{eq:virial}
\end{align}
where $\mh=\mu_{B}/T$ is the ratio of baryon chemical potential over temperature and $p_{k}$ are coefficients to be determined. Therefore, the partition
function, $Z$, can be written as
\begin{align}
  Z=\exp\left[ VT^{3} \sum_{k=0}^{\infty}p_{k}(T)\cosh\left( k \mh \right) \right].
  \label{eq:Z_1}
\end{align}
On the other hand the partition function can also be written as a sum involving  all the
$\dN$ net-baryon partition functions, $z_{\dN}$, $-\infty \leq \dN \leq \infty$ 
\begin{align}
Z=\sum_{\dN=-\infty}^{\infty}z_{\dN}e^{\dN \mh} = z_{0} + 2 \sum_{\dN=1}^{\infty} z_{\dN}\cosh(\dN \mh).
  \label{eq:Z_2}
\end{align}
Here and in the following we  denote the net-baryon number by $\dN=n_{B}-n_{\bar{B}}$. In the previous equation we 
made use of charge symmetry to relate $z_{\dN}=z_{-\dN}$. In the present paper we are interested in the case of vanishing baryon chemical potential, where the cumulants are measured on the lattice up to the eighth order. Given the $z_{\dN}$, the probability to have
$\dN$ net-baryons at vanishing chemical potential, $\mh=0$, is
then given by
\begin{align}
  P(\dN)= \frac{z_{\dN}}{Z(\mh=0)} .
  \label{eq:P_n}
\end{align}
Therefore, the task at hand is to determine the $\dN$ net-baryon partition function, $z_\dN$. To this end we
equate the two expressions for the partition function, Eqs.~(\ref{eq:Z_1}) and (\ref{eq:Z_2}), and divide
both sides by the partition function taken at $\mh=0$,
\begin{eqnarray}
\frac{\exp \left[ VT^{3}\sum_{k=1}^{\infty }p_{k}(T)\cosh \left( k\hat{\mu}%
_{B}\right) \right] }{\exp \left[ VT^{3}\sum_{k=1}^{\infty }p_{k}(T)\right] }
&=&\frac{1}{Z(\hat{\mu}_{B}=0)}\left( z_{0}+2\sum_{\mathcal{N}=1}^{\infty
}z_{\mathcal{N}}\cosh (\mathcal{N}\hat{\mu}_{B})\right)   \notag \\
&=&P(0)+2\sum_{\mathcal{N}=1}^{\infty }P(\mathcal{N})\cosh (\mathcal{N}\hat{%
\mu}_{B}) .
\end{eqnarray}%
We note that the normalization with $Z(\mh=0)$ removes the dependence of the left hand side on the virial
coefficient $p_{0}$ and thus the sums start at $k=1$. 
Upon rotating to imaginary chemical potential (see also \cite{Morita:2013tu}), $\mh\rightarrow i \mb$, the above equation turns into
\begin{align}
  \frac{\exp\left[ VT^{3} \sum_{k=1}^{\infty}p_{k}(T)\cos\left( k \mb \right) \right]}{\exp\left[
  VT^{3} \sum_{k=1}^{\infty}p_{k}(T)\right]} = P(0) + 2 \sum_{\dN=1}^{\infty} P(\dN)\cos(\dN \mb) ,
  \label{}
\end{align}
so that the $\dN$-net-baryon probabilities, $P(\dN)$, result from simple Fourier transform 
\begin{align}
  P(\dN)=\frac{1}{\pi}\int_{0}^{\pi}d\mb \, 
  \cos(\dN \mb)\,  \frac{\exp\left[ VT^{3} \sum_{k=1}^{\infty}p_{k}(T)\cos\left( k \mb \right) \right]}{\exp\left[
  VT^{3} \sum_{k=1}^{\infty}p_{k}(T)\right]} .
  \label{eq:p_from_fourier}
\end{align}
Therefore, given the coefficients $p_{k}(T)$, one can determine the multiplicity distribution
$P(\dN)$. Furthermore, since the cumulants are given by
\begin{align}
  \cum_{n}=VT^{3}\frac{\partial^{n}}{\partial\mh^{n}}\frac{P}{T^{4}} ,
  \label{eq:cum_from_P}
\end{align}
they can be readily calculated either using Eq.~\eqref{eq:virial} or directly from the multiplicity
distribution, $P(\dN)$. This provides an important cross check for the numerical Fourier transform required.
 
In order to proceed, all we need then are the coefficients,
$p_{k}$. Here, we make use of the recent work by Vovchenko et al. \cite{Vovchenko:2017gkg}. In
this paper the authors used a cluster model to determine the coefficients $b_{k}(T) = k \, p_{k}(T)$
provided  the first two coefficients, $b_{1}(T)$ and $b_{2}(T)$ are known. In \cite{Vovchenko:2017xad} the first
four coefficients, $b_{1}(T)\ldots b_{4}(T)$ have been determined by lattice QCD methods. Taking the
lattice results for  $b_{1}(T)$ and $b_{2}(T)$ as input, the cluster model of \cite{Vovchenko:2017gkg} not only
reproduces $b_{3}(T)$ and $b_{4}(T)$ obtained from lattice QCD, but also is able to reproduce the
baryon number cumulants obtained from lattice QCD up to eighth order, which is the highest presently available
(for the quality of the agreement see \cite{Vovchenko:2017gkg,Vovchenko:2018zgt}).
Given the success of this cluster model, we consider it to be a suitable  model to
determine a more or less realistic net-baryon multiplicity distribution. Also, alternative virial expansion of
Ref.~\cite{Almasi:2018lok} leads to nearly identical results.  We further  note, that model
dependence may be systematically reduced by extracting ever higher coefficients $b_{k}(T)$ from
lattice QCD. Finally, we recall from Refs. \cite{Vovchenko:2017gkg,Vovchenko:2018zgt}, that in their cluster expansion model the
above sums can be carried out analytically, which is helpful for carrying out the  Fourier transform.
\begin{align}
 \sum_{k=1}^{\infty}p_{k}(T)\cosh\left( k \mh \right) = 
\frac{b_{1}^{2}\left( 3+16\pi ^{2}\right) }{%
b_{2}16\left( 3+4\pi ^{2}\right) ^{2}}\left[ 4\pi ^{2}\text{Li}%
_{2}(f_{\mh })+4\pi ^{2}\text{Li}_{2}(f_{-\mh })+3\text{Li}%
_{4}(f_{\mh })+3\text{Li}_{4}(f_{-\mh })\right] ,
  \label{eq:sum1}
\end{align}
where
\begin{equation}
f_{\mh}=\frac{8b_{2}\left( 3+4\pi ^{2}\right) }{b_{1}\left(
    3+16\pi ^{2}\right) } e^{\mh };\quad \mh =\frac{\mu _{B}}{T}; \quad b_{k} = k\, p_{k} ,
\label{eq:fmub}
\end{equation}
and $\text{Li}_{n}(z)$ is the polylogarithm.\footnote{An analogous sum with $\cosh\left( k \mh \right) \rightarrow \cos\left( k \mb \right)$ can by obtained by $\mh\rightarrow i \mb$.}

\begin{figure}[t]
\begin{center}
\includegraphics[scale=0.36]{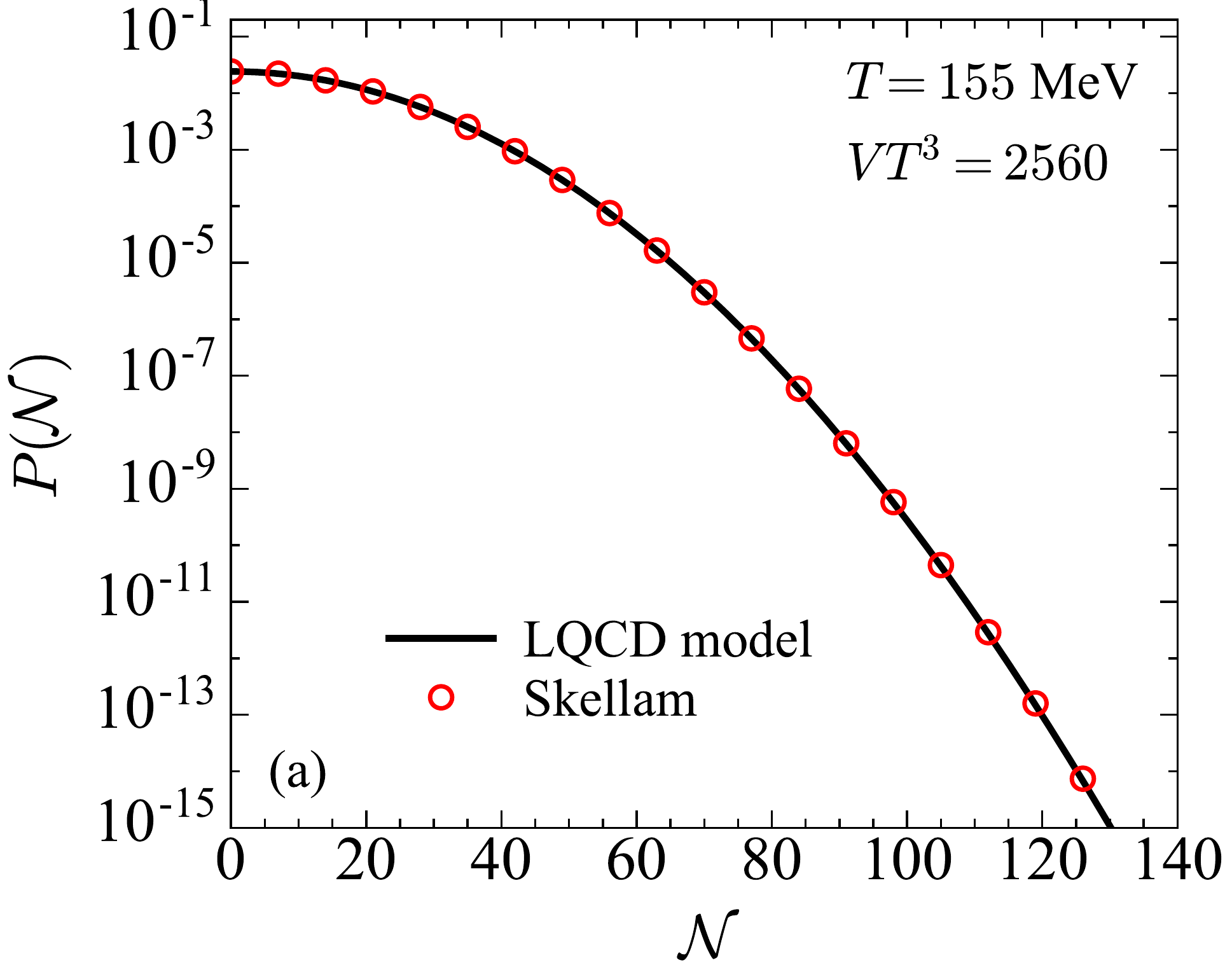}\hspace{6mm} 
\includegraphics[scale=0.36]{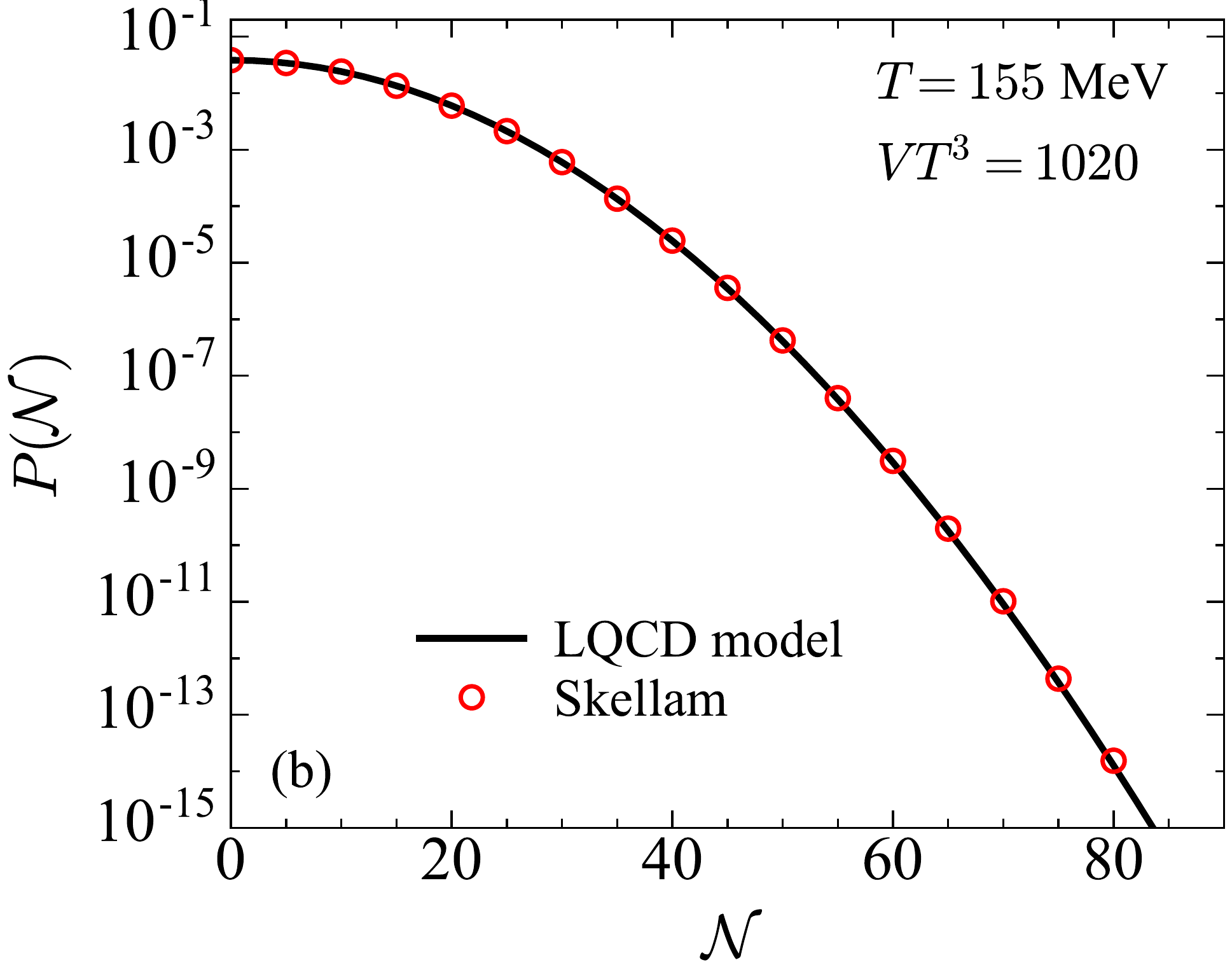}
\end{center}
\par
\vspace{-5mm}
\caption{LQCD-based net-baryon multiplicty distribution at $T=155$ MeV compared with the Skellam distrubution at the same 
width for (a) $V=5280$ fm$^3$ and (b) $V=2100$ fm$^3$. For clarity we only show a limited number of
points for the Skellam distribution. The results are based on the cluster expansion  model of Ref.~\cite{Vovchenko:2017gkg} and are consistent with all net-baryon cumulants presently known from lattice QCD.}
\label{fig:155-multdist}
\begin{center}
\includegraphics[scale=0.36]{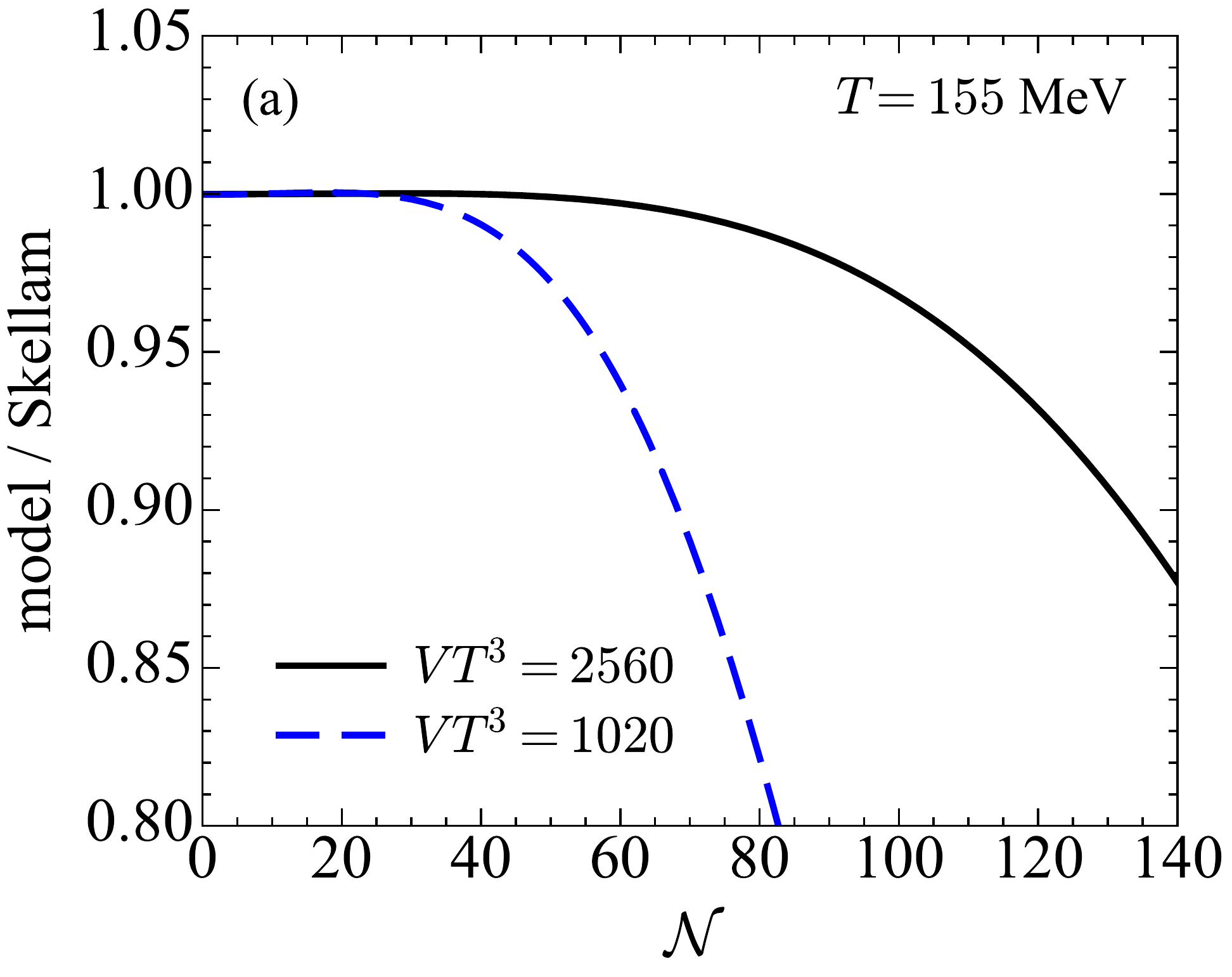}\hspace{6mm} 
\includegraphics[scale=0.36]{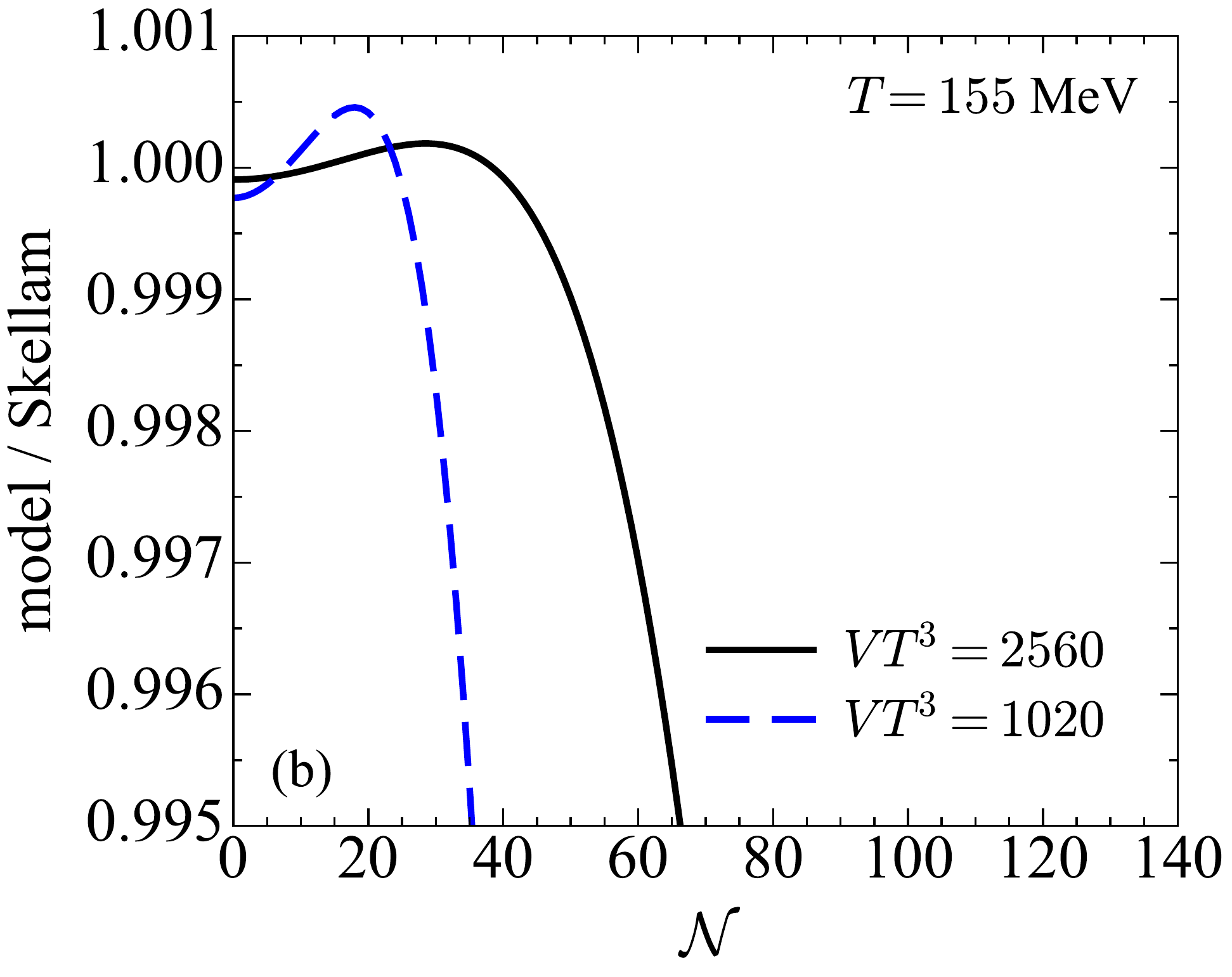}
\end{center}
\par
\vspace{-5mm}
\caption{The ratios (model over Skellam) of the distributions presented in Fig. \ref{fig:155-multdist} 
for (a) broad range of $\dN$ and (b) close to $\dN \sim 0$.}
\label{fig:155-ratio-all}
\end{figure}

\section{Results}
\label{sec:Results}

\begin{figure}[t]
\begin{center}
\includegraphics[scale=0.36]{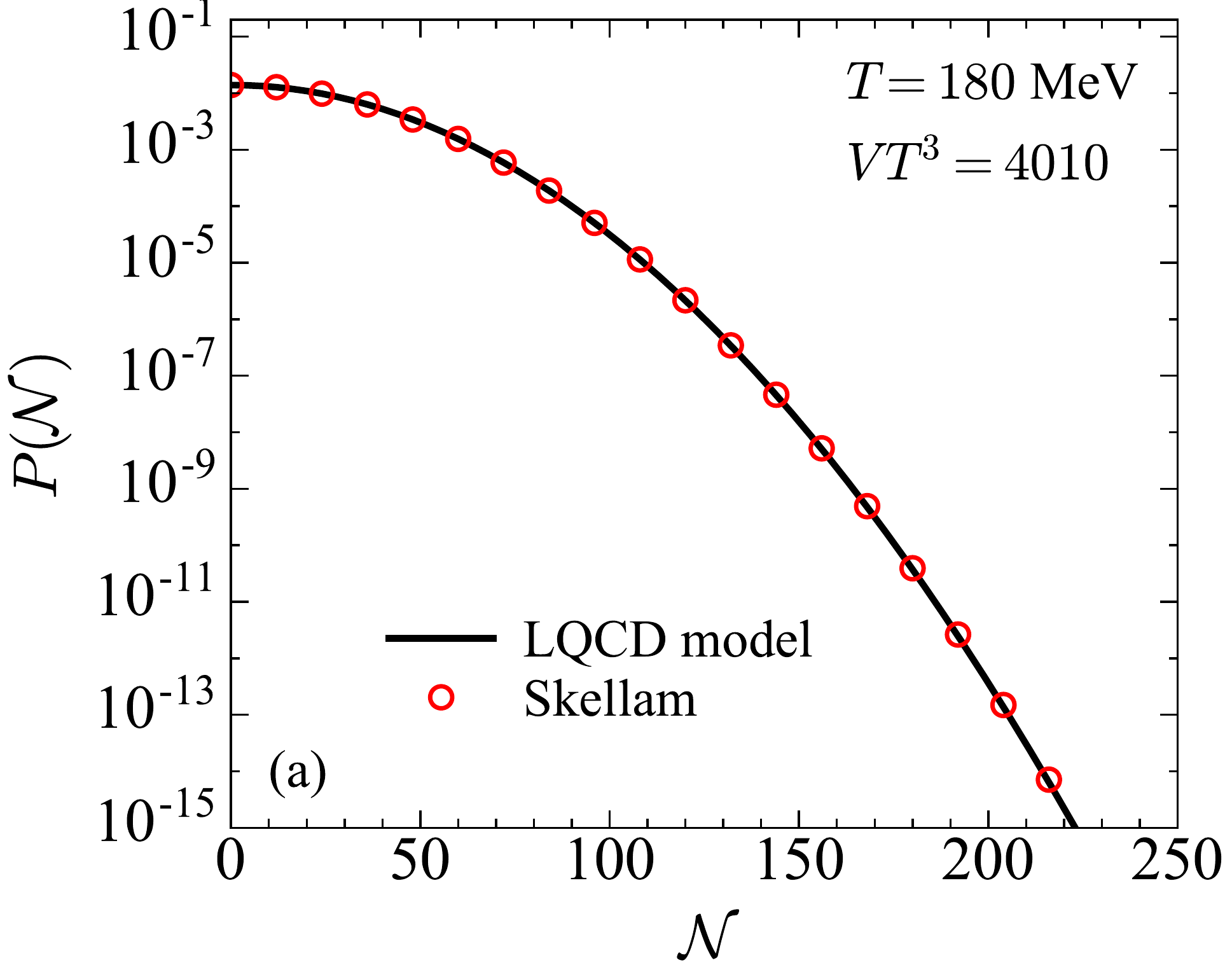}\hspace{6mm} 
\includegraphics[scale=0.36]{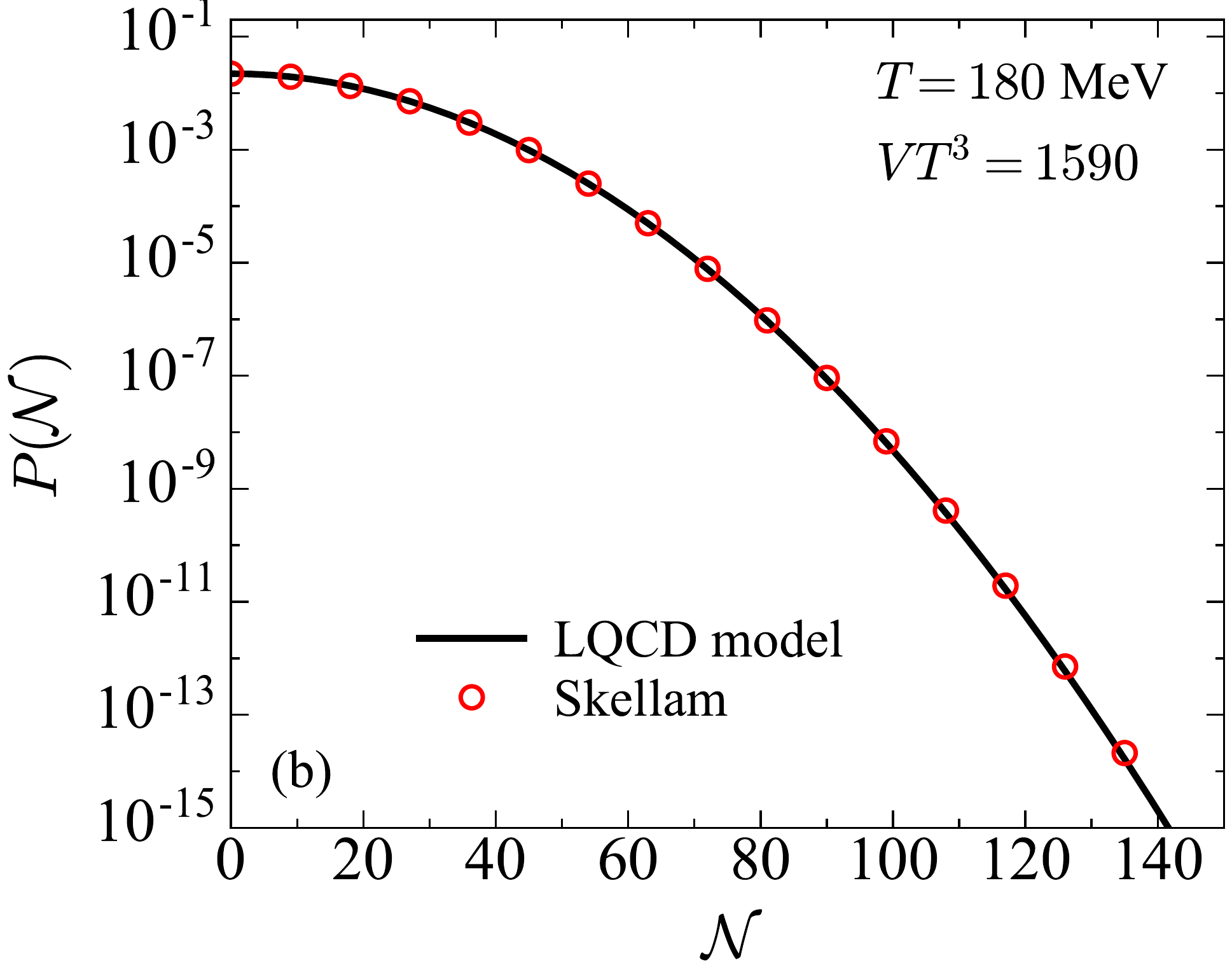}
\end{center}
\par
\vspace{-5mm}
\caption{LQCD-based net-baryon multiplicty distribution at $T=180$ MeV compared with the Skellam distrubution at the same 
width for (a) $V=5280$ fm$^3$ and (b) $V=2100$ fm$^3$. For clarity we only show a limited number of
points for the Skellam distribution. The results are based on the cluster expansion  model of Ref.~\cite{Vovchenko:2017gkg} and are consistent with all net-baryon cumulants presently known from lattice QCD.}
\label{fig:180-multdist}
\begin{center}
\includegraphics[scale=0.36]{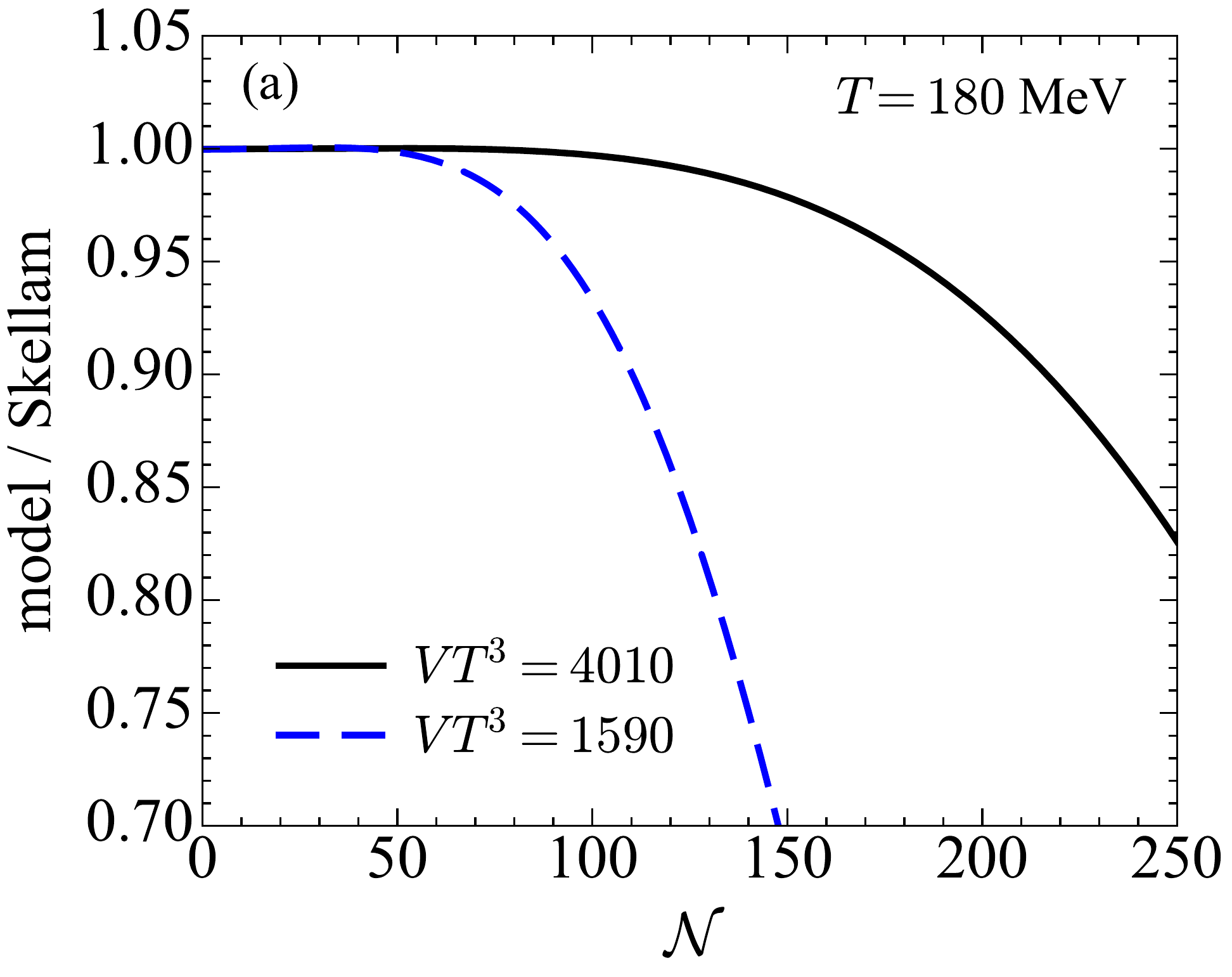}\hspace{6mm} 
\includegraphics[scale=0.36]{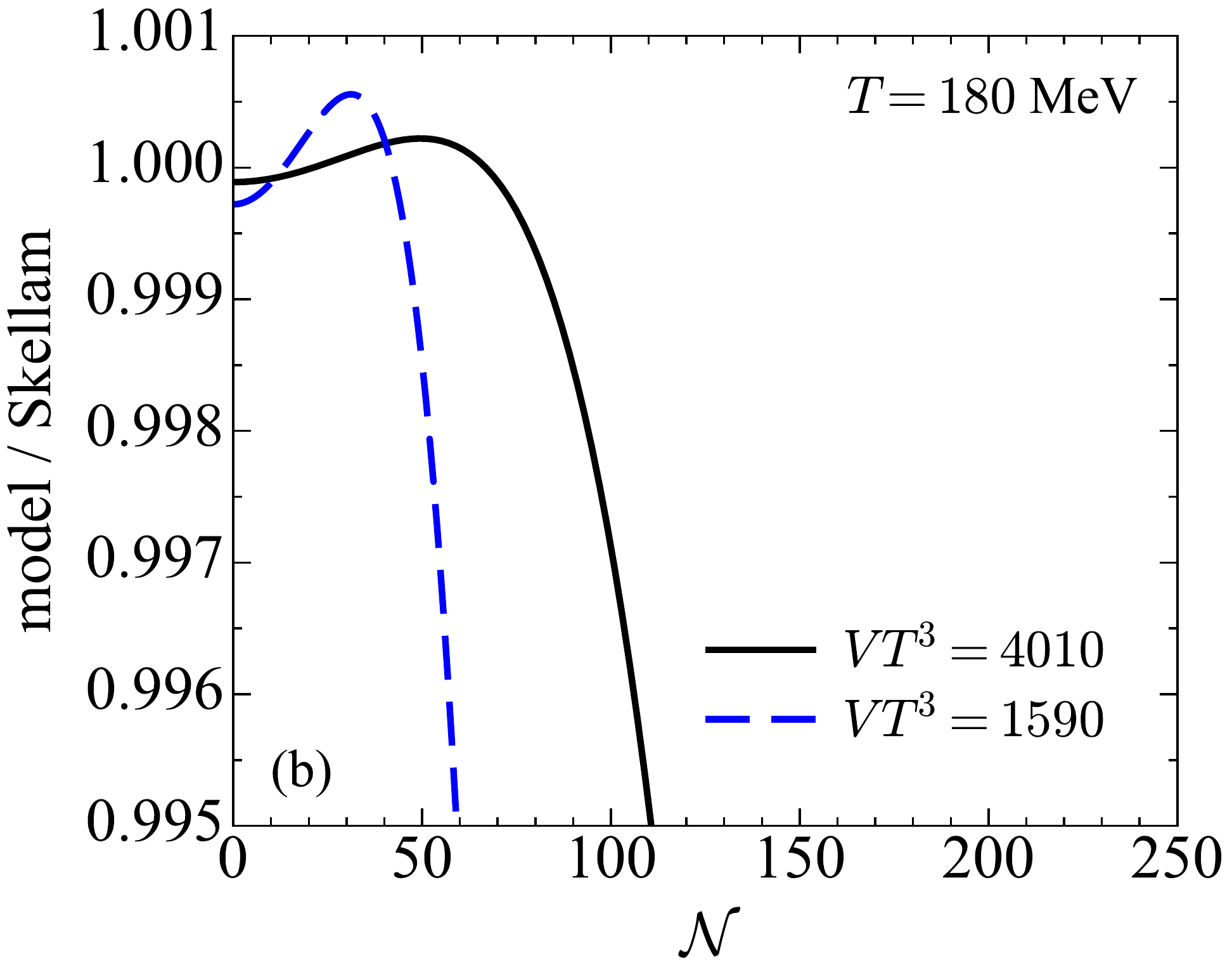}
\end{center}
\par
\vspace{-5mm}
\caption{The ratios (model over Skellam) of the distributions presented in Fig. \ref{fig:180-multdist} for (a) broad range 
of $\dN$ and (b) close to $\dN \sim 0$.}
\label{fig:180-ratio-all}
\end{figure} 

\begin{figure}[t]
\begin{center}
\includegraphics[scale=0.36]{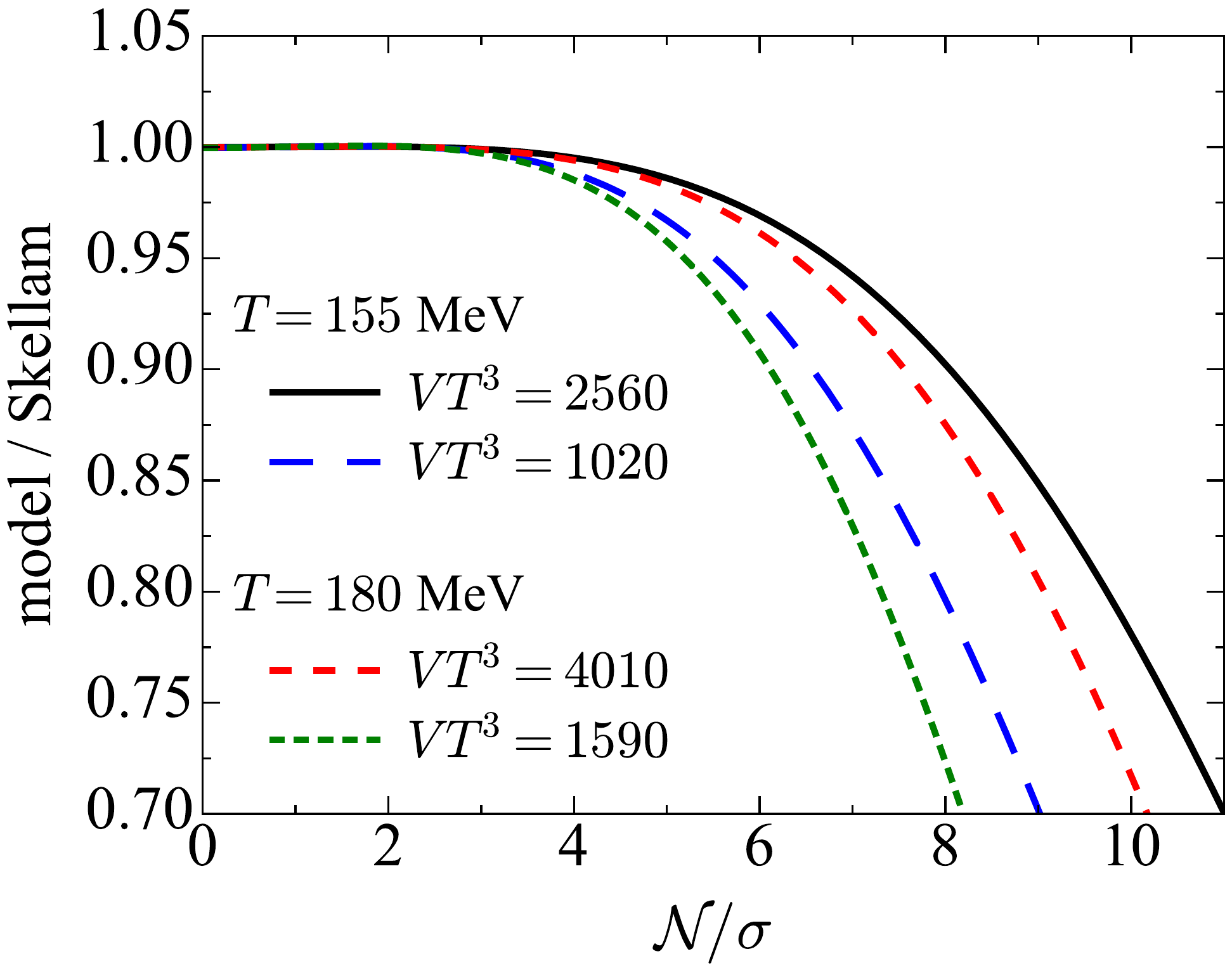}
\end{center}
\par
\vspace{-5mm}
\caption{The ratios of the distributions presented in Figs. \ref{fig:155-multdist} and
  \ref{fig:180-multdist} with $\dN$ rescaled by the respective widths, $\sigma$ (see Eq.~\eqref{eq:sigma}). }
\label{fig:ratio_all}
\end{figure} 

After having introduced the method for extracting the multiplicity distribution $P(\dN)$ let us now turn
to the results. Given the cluster model of \cite{Vovchenko:2017gkg} the kernel of the Fourier integral of
Eq.~\eqref{eq:p_from_fourier} is straightforward to determine (see Eqs. \eqref{eq:sum1} and \eqref{eq:fmub}). 
We further note that the kernel of the Fourier integral depends on the volume so that with increasing volume
it becomes an ever steeper function localized around $\mb=0$. As a result, with increasing $V$ ever
higher Fourier components become relevant. This of course simply reflects the fact that for fixed
temperature the average number of baryons and anti-baryons increases with the volume resulting in an ever
broader distribution of the  net-baryon number. In the following, we will consider two volumes,
$V_{1}=5280 \,\rm fm^{3}$ which corresponds to the chemical freeze out volume extracted \cite{Andronic:2017pug} from data
taken by the ALICE collaboration at the LHC and $V_{2}=2100\,\rm fm^{3}$ which is the estimated chemical freeze out
volume for the top RHIC energy ($\sqrt{s}=200\gev$) \cite{Andronic:2014zha}. 
For the multiplicity distributions shown in the following we have ensured that the maximum value for $\dN$,
$\dN_{\rm max}$, is such that all cumulants up to
$12^{th}$ order agree within one percent if calculated either from Eq.~\eqref{eq:cum_from_P} or in terms of
moments of the multiplicity distribution, $\ave{\dN^{k}}=\sum_{\dN=-\dN_{\rm max}}^{\dN_{\rm max}}P(\dN) \dN^{k}$. We will consider two
values for the temperature. The lower one, $T=155\mev$, corresponds the chemical freeze-out temperature extracted from hadron-yield
systematics \cite{Andronic:2017pug}. We also consider
$T=180\mev$, where the cumulants are substantially different. For example $\cum_{4}/\cum_{2}=0.725\pm0.0529$ at
$T=155\mev$ while $\cum_{4}/\cum_{2}=0.273\pm0.0189$ at $T=180\mev$ \cite{Bazavov:2017dus}. This will 
give us an idea to which extent the multiplicity
distribution depends on the value of the cumulants. The values for the input parameters $b_{1},\,b_{2}$ are  
$b_{1}= 0.113907$, $b_{2}=-0.00428816$ for $T=155\mev$ and 
$b_{1}=0.298302$, $b_{2}=-0.0697688$ for $T=180\mev$.

In Fig. \ref{fig:155-multdist} we show the multiplicity distribution, $P(\dN)$, for a system at temperature
$T=155\mev$. Panel (a) corresponds to the volume extracted for LHC and panel (b) for that at the top RHIC energy,
corresponding to $V_{\rm LHC}T^{3}=2560$ and $V_{\rm RHIC}T^{3}=1020$, respectively. Note that we show the distributions
only for positive $\dN$ since it is symmetric with respect to $\dN \Leftrightarrow -\dN$.
In addition, as red circles we show the corresponding Skellam distribution
\begin{align}
P_{\rm Skellam}(\dN)= e^{-2 \Lambda} I_{\dN}(2 \Lambda) ,
  \label{eq:skellam}
\end{align}
where $I_{\dN}$ is the modified Bessel function of the first kind. 
We adjust the parameter $\Lambda$ such that the Skellam distribution and the model distribution
have the same variance, 
\begin{align}
\sigma^{2}= \cum_{2} = \ave{\dN^2},
  \label{eq:sigma}
\end{align}
that is $\Lambda = \sigma^{2}/2$ since for Skellam $\ave{\dN^2}=2 \Lambda$ (of course $\ave{\dN}=0$). Here
$\cum_{2}$ is the second order cumulant, consistent with lattice QCD. The difference between the model
distribution, $P(\dN)$, which is consistent with the cumulants from lattice QCD, and the Skellam distribution, where
all cumulants are the same, $\cum_{2 n}=\cum_{2}$, is barely visible. To  show the difference
more clearly, in Fig. \ref{fig:155-ratio-all} we show the ratio of $P(\dN)/P_{\rm Skellam}(\dN)$. 
In panel (a) we plot the ratio over the same range as in the previous figure, 
and we see for the smaller volume this ratio drops faster. In panel (b) we zoom 
into the region where the ratio is approximately unity. We see that in both cases the ratio
exceeds one, however, only by less than 0.1\%. Also, the maximum in case of the RHIC volume is  more
pronounced. To see how things change with temperature, we show the resulting multiplicity
distributions also for a temperature of $T=180\mev$ in Figs.~\ref{fig:180-multdist} and
\ref{fig:180-ratio-all}. We used the same volumes used in Figs.~\ref{fig:155-multdist} and
\ref{fig:155-ratio-all} so that the factor $VT^{3}$ and thus the width of the distribution
increase with the temperature.  Although the cumulants are quite different at $T=180\mev$, comparing
Figs.~\ref{fig:155-ratio-all} and \ref{fig:180-ratio-all} 
we find that the deviation from the Skellam distribution does not change significantly.
For both temperatures we have observed a volume dependence of the multiplicity distributions, which
qualitatively is not surprising, as the width increases with the volume $\sigma^{2}=\cum_{2}\sim
V$. To remove this rather trivial effect, in Fig.~\ref{fig:ratio_all} 
 we plot the ratios $P(\dN)/P_{\rm Skellam}(\dN)$ as a
function of $\dN/\sigma$, where $\sigma$ is respective width for each  volume and
temperature. Obviously the distribution does not simply scale with the width.

\begin{table}
\begin{tabular}{|l|l|l|l|} \hline
       & $\cum_{4}/\cum_{2}$ & $\cum_{6}/\cum_{2}$ & $\cum_{8}/\cum_{2}$ \\ \hline
$T=155,\, VT^{3}=2560$, (Fig. 1(a)) & 86  ($3.1\times10^{-8}$) & 118  ($2.3\times10^{-13}$)    & 137 ($3.8\times10^{-17}$)  \\ \hline
$T=155,\, VT^{3}=1020$, (Fig. 1(b)) & 52  ($1.6\times10^{-7}$) & 72   ($2.6\times10^{-12}$)    & 83  ($1.5\times10^{-15}$)  \\ \hline
$T=180,\, VT^{3}=4010$, (Fig. 3(a)) & 161 ($1.9\times10^{-9}$) & 211  ($2.3\times10^{-14}$)    & 249 ($5.7\times10^{-19}$)  \\ \hline
$T=180,\, VT^{3}=1590$, (Fig. 3(b)) & 98  ($8.7\times10^{-9}$) & 128  ($2.7\times10^{-13}$)    & 150 ($2.4\times10^{-17}$)   \\ \hline
\end{tabular}
\caption{Value of the minimum number of net-baryon $\dN_{\rm min}$ one needs to sum $P(\dN)$ over in
order to obtain $\cum_{n}/\cum_{2}$ within $5\%$ or less. 
If we demand $10\%$ or less the numbers are reduced by $\dN \sim 3$. We also give the values of
$P(\dN_{\rm min})$ in parenthesis.} 
\label{table:t1}
\end{table}

\section{Discussion and Conclusion}

After having presented the multiplicity distributions which are consistent with the cumulants
determined from lattice QCD a few points are worth discussing

\begin{enumerate}[(i)]
\item Our main finding is that the deviation of the LQCD-based net-baryon multiplicity distribution from the Skellam
  distribution with the same width (or same second order cumulant) is very small for all cases
  considered. For the more realistic temperature of $T=155\mev$ and a value of the probability as small as
  $P(\dN)\simeq 10^{-15}$ the difference is  less than 15\% for LHC energies and at best 20\% in case of
  top RHIC energies
  (see Figs.~\ref{fig:155-multdist} and  \ref{fig:155-ratio-all}). The same qualitative result is 
  found when using the virial
  expansion of Ref.~\cite{Almasi:2018lok}, which also reproduces the lattice cumulants within errors
  but has different asymptotic behavior of the virial coefficients, $p_{k}(T)$. In this case, as
  shown in Fig.~\ref{fig:ratio_friman}, the ratio to the corresponding Skellam distribution is
  somewhat closer to unity as  compared to the model of
  \cite{Vovchenko:2017gkg} (see Fig.~\ref{fig:155-ratio-all}(a)), while the  multiplicity
  distributions look virtually identical to those shown in
  Fig.~\ref{fig:155-multdist}. Specifically, the deviations are  $\sim 5\%$ for LHC and $\sim 10\%$
  for RHIC at the points where $P(\dN)\simeq 10^{-15}$. 
  Therefore, given the small deviation from a Skellam distribution and the fact 
  that the direct measurement of a multiplicity
  distribution involves unfolding of efficiency corrections (see, e.g., \cite{Bzdak:2016qdc}) it is
  very unlikely that a meaningful extraction of the  multiplicity distribution is possible in
  practice.
  
\begin{figure}[t]
\begin{center}
\includegraphics[scale=0.36]{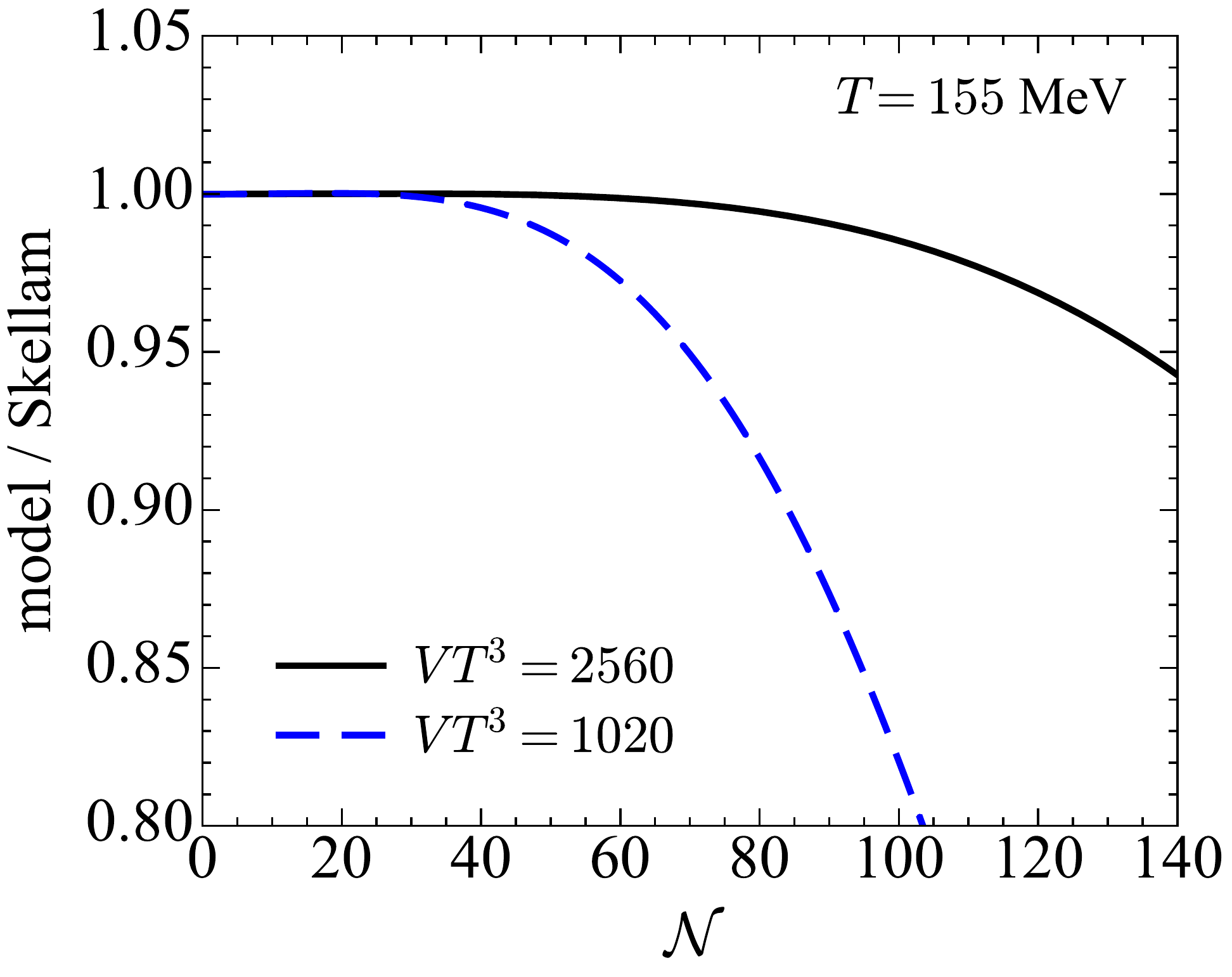}
\end{center}
\par
\vspace{-5mm}
\caption{The ratios (model over Skellam) for the virial expansion model of Ref. \cite{Almasi:2018lok}.}  
\label{fig:ratio_friman}
\end{figure} 

\item
  Here we have discussed the multiplicity distribution of net-baryons. In experiment one is
  usually restricted to the measurement of net-protons. As discussed in \cite{Kitazawa:2011wh},
  assuming fast isospin exchange the net-proton distribution can be derived from the net-baryon
  distribution by folding with a binomial distribution with Bernoulli probability of $p\simeq
  0.5$. This, however, brings the distribution even closer to a Skellam distribution \cite{Bzdak:2012ab}.

\item
  In order to provide some guidance how sensitive the cumulants are to the tails of the distribution,
  in Table~\ref{table:t1} we provide the minimum
  range $\dN_{\rm min}$ one has to sum the  multiplicity distribution over in order to get a certain cumulant 
  ratio within 5\%  of the correct value ($\dN_{\rm min}$ is reduced by 3 if for 10\% accuracy).
  We also list the probability $P(\dN_{\rm min})$ at this point. For example, in order to obtain
  $\cum_{6}/\cum_{2}$ within 5\% for the LHC at the realistic freeze-out temperature of 
  $T=155\mev$ ($VT^{3}=2560$), we need to sum until $\dN=118$, where the probability is as low as $10^{-13}$.
  
\item
  To estimate the required statistics for the measurement of a given cumulant, one either samples
  the multiplicity distribution or equivalently makes use the delta method
  \cite{davison2003statistical,Luo:2014rea,Luo:2017faz}. This is discussed in some  detail in the Appendix.
  Using the delta method, one finds that the relative error
  for the measurement of a cumulant of given order $n$, $K_{n}$, in principle depends on all
  cumulants up to order $2 n$. However, as shown in the Appendix, in case of QCD where the cumulant
  ratios $K_{n}/K_{2}$ are of order one, the leading term involving only the second order cumulant
  dominates the {\em absolute} error, $\Delta K_{n}$. The {\em relative } error, $\Delta
  K_{n}/\left|K_{n}\right|$ of course depends on the actual magnitude of $K_{n}$. As shown in
  detail in the Appendix, to leading order the relative error the fourth and sixth order cumulants are  given by:
  \begin{align}
  \frac{\Delta K_{4}}{\left|K_{4}\right| } & \simeq\frac{\sqrt{24} }{\sqrt{\neve} }
    \frac{K_{2}}{\left| R_{4,2}\right|},\\
  \frac{\Delta K_{6}}{\left|K_{6}\right| } & \simeq\frac{\sqrt{720} }{\sqrt{\neve} }
    \frac{K_{2}^{2}}{\left| R_{6,2} \right|},
  \end{align}
where $R_{n,2}=K_{n}/K_{2}$ denotes the cumulant ratio of the n-th order cumulant over the second order cumulant.
Given the relative error, one can estimate the number of events required to measure the cumulants and, as
discussed in the Appendix, we find that for the LHC conditions at least $2.7 \times 10^{8}$ 
and $5.3 \times 10^{15}$  events are needed to measure $K_{4}$ and $K_{6}$ at 10\% accuracy, respectively. For RHIC
one needs at least  $4.4 \times 10^{7} $ and  $1.4 \times 10^{14}$. As further discussed in the
Appendix, the situation is somewhat less statistics hungry in case of net protons, simply because the
value for the second order cumulant, $K_{2}^{p}$ is smaller in this case. Assuming that the 
cumulant ratios are the same for net-protons as for net-baryons,
$4.8\times 10^{6}$ and $ 1.8\times 10^{12}$ events are required for $K_4^p$
and $K_{6}^p$, respectively in case of LHC conditions. The corresponding values for RHIC are $2.5
\times 10^{6}$ and $4.9\times 10^{11}$ events. Assuming that the cumulant ratios are close to $1$ 
for the net-protons, these
values reduce to $3.2 \times 10^{6}$ and $1.3 \times 10^{11}$ events for $K_{4}^{p}$ and $K_{6}^{p}$ at the LHC. 
The corresponding number of events for RHIC are $1.7\times 10^{6}$ and $ 3.8 \times 10^{10}$.
 
\item We note that, although the  multiplicity distributions for the same temperature but different volumes are
  different, they are mathematically related. Using the fact that all cumulants scale with the volume, 
  $\cum_{n}\sim V$, one
  can easily show that multiplicity distribution at one volume, $P(\dN,V_2)$, is related to that at another volume,
  $P(\dN,V_1)$ via\footnote{Assuming that all cumulants scale with the volume is equivalent to the assumption that the cumulnat generating function, $K(t;V)=\ln(\sum_{\dN} P(\dN;V) e^{\dN t})$ is proportional to the volume, i.e, $K(t;V) \sim V$. Consequently $K(t;V_{2})/K(t;V_{1})=V_{2}/V_{1}$, which can be written as $e^{K(t;V_{2})}=(e^{K(t;V_{1})})^{V_{2}/V_{1}}$. Using the definition of $K(t)$, changing the variable $t\rightarrow it$ and using $\delta _{n,m}=\frac{1}{2\pi }\int_{0}^{2\pi }e^{i(n-m)t}dt$, we obtain Eq. (\ref{eq:P_relation}).} 
  \begin{align}
    P(\dN;V_{2})=\frac{1}{2\pi}\int_{0}^{2\pi} dt \left[ \sum_{\dM=-\infty}^{\infty} 
    P(\dM ;V_{1}) e^{i\dM t}\right]^{V_{2}/V_{1}}e^{-i \dN t} .
   \label{eq:P_relation}
  \end{align}
  In the case of $V_{2}/V_{1}=2$ this simplifies to 
  \begin{align}
    P(\dN;V_{2})= \sum_{\dM=-\infty}^{\infty} P(\dM ;V_{1}) P(\dN - \dM;V_{1}) .
   \label{eq:P_relation_2}
  \end{align}
  
\item The results we presented here used the central values for input parameters, $b_{1}$ and
  $b_{2}$. Taking the errors into account does not change the conclusions of this paper. For example, for
  $T=155\mev$ the error on $b_{2}$ is quite sizable, $\sim 30\%$ (it is only $\sim 2\%$ at
  $T=180\mev$) \cite{Vovchenko:2017xad}.  The resulting uncertainty on the ratio
  $P(\dN)/P_{\rm Skellam}(\dN)$, is negligible for small $\dN$ and of the order of $\sim \pm 4\%$ for
  $\dN\simeq 140$ in case of the LHC volume, $VT^{3}=2560$. 
  
\item Of course our result is model dependent in the sense that it relies on the cluster expansion model of
  \cite{Vovchenko:2017gkg}. However, this model reproduces, within errors, all the cumulants so far calculated on the
  lattice. Therefore, we consider our results for the multiplicity distributions reasonably realistic. 
  Also, the model dependence can be reduced by calculating higher order virial coefficients $b_{n}$ on the lattice.
  
\item We note that the above formalism can be readily extended to finite values of the baryon-number 
  chemical potential $\mu_{B}$. However, with increasing  $\mu_{B}$ the resulting multiplicity distribution  
  $P(\dN)$  will be ever more sensitive to the knowledge of higher order virial coefficients, $p_{k}$, which 
  are not yet constrained by lattice QCD. This would increase the model dependence and,
  therefore, we restricted ourselves to the case of $\mu_{B} = 0$. 
\end{enumerate}

In conclusion, utilizing the cluster expansion model of \cite{Vovchenko:2017gkg} we have derived a
multiplicity distribution of net-baryons which is consistent with the net-baryon number cumulants
determined from lattice QCD. We find that the resulting distributions are very close to a Skellam
distribution with the same width.
We further applied the delta method to calculate the expected statistical error for a measurement 
of fourth and sixth order cumulants and based on this, estimated the required number of events.

\acknowledgments
We would like to thank V. Vovchenko for discussions about the cluster expansion model and for providing us with the
lattice results for $b_{1}$ and $b_{2}$. We further want to thank B. Friman and K. Redlich for
discussions concerning their virial expansion approach. We thank the Hot-QCD and Wuppertal-Budapest collaborations
for providing us with
the results for the cumulant ratios. 
This work was stimulated  by the discussions at the Terzolas meeting on “Heavy-Ion Physics in the 2020’s”
(Terzolas, Italy, May 19-21, 2018). 
A.B. is partially supported by the
Faculty of Physics and Applied Computer Science AGH UST statutory tasks No. 11.11.220.01/1 within
subsidy of Ministry of Science and Higher Education, and by the National Science Centre, Grant
No. DEC-2014/15/B/ST2/00175.  V.K. was supported by the U.S. Department of Energy, Office
of Science, Office of Nuclear Physics, under contract number DE-AC02-05CH11231.  This work also
received support within the framework of the Beam Energy Scan Theory (BEST) Topical Collaboration.

\appendix
\section{Estimating the required statistics using the delta method}

In order to estimate the needed statistics, one needs to know the expected error given a certain
number of events. This error can either be determined by sampling the multiplicity distribution many
times or via the delta method (see, e.g., \cite{davison2003statistical,Luo:2014rea,Luo:2017faz} for
details). In Refs. \cite{Luo:2014rea,Bzdak:2018axe} it has be demonstrated  that both methods lead
to identical results. Since sampling the multiplicity distribution shown in
Fig.~\ref{fig:155-multdist} is numerically very demanding, here we apply the delta method. We
are interested in the expected error of the cumulants $K_{n}$ in which case the application of
the delta method is straightforward.  The
random variables are the moments about zero, $\mu_{k}=\ave{N^{k}}$. Therefore, we express the
cumulant, $K_{n}$, in terms of the moments, $K_{n}=F_{n}(\mu_{1}, \ldots, \mu_{n})$. Then
according to the delta method the variance of $K_{n}$ for a sample with $\neve$ events is given by
\begin{eqnarray}
  \mathrm{Var}\left( K_{n}\right)  &=&\sum_{i=1}^{n}\sum_{j=1}^{n}\frac{%
  \partial F_{n}}{\partial \mu _{i}}\frac{\partial F_{n}}{\partial \mu _{j}}\mathrm{Cov%
  }\left( \mu _{i},\mu _{j}\right) ,  \label{eq:delta} \\
  \mathrm{Cov}(\mu _{i},\mu _{j}) &=&\frac{1}{n_{\mathrm{events}}}\left( \mu
  _{i+j}-\mu _{i}\mu _{j}\right) .
\end{eqnarray}
The {\em absolute} error is then  given by $\Delta K_{n} = \sqrt{\mathrm{Var}\left( K_{n} \right) }$.
After re-expressing the moments $\mu_{i}$ in terms of the cumulants, $K_n$, we obtain for the
variance of the fourth order cumulant 
\begin{align}
  \mathrm{Var}\left(  K_{4} \right) 
  &=\frac{1}{\neve } \left( 24 K_2^4+72 K_4 K_2^2+16 K_6 K_2+34 K_4^2+K_8 \right)
  \non
  &=\frac{1}{\neve } 24 K_2^4 \left(1+\frac{3 R_{4,2}}{K_2}+\frac{17 R_{4,2}^2}{12 K_2^2}+\frac{2 R_{6,2}}{3 K_2^2}+\frac{R_{8,2}}{24 K_2^3}\right) .
  \label{eq:k4_Var}
\end{align}
Here we used the fact that the cumulants of odd order, $K_{1},K_{3}\ldots$ vanish at $\mb=0$. Also,
in the second line we have introduced the cumulant ratios, $R_{n,2}=K_{n}/K_{2}$ and factored
out the leading term. The relative error, $\Delta K_{4}/K_{4}$ is then given by
\begin{align}
  \frac{\Delta K_{4}}{\left|K_{4}\right|} =\frac{\sqrt{24} }{\sqrt{\neve} }
  \frac{K_{2}}{\left|R_{4,2}\right|}\sqrt{ 1+\frac{3 R_{4,2}}{K_2}+\frac{17 R_{4,2}^2}{12 K_2^2}
  +\frac{2 R_{6,2}}{3 K_2^2}+\frac{R_{8,2}}{24 K_2^3} }  \,.
  \label{eq:k4_error}
\end{align}

We note, that the above expressions depend only on the cumulant ratios, $R_{n,2}$, and the second order
cumulant, $K_{2}$. The former can be, and have been up to eighth order \cite{Borsanyi:2018grb},
determined by lattice QCD. The second order cumulant, $K_2$, on the other hand requires additional experimental
input, namely the temperature and volume of the system, since $K_{2}=V T^{3}\chi_{2}$ with
$\chi_{2}$ being the second order susceptibility, which has been determined on the lattice.
Using the values for  $VT^{3}=2560$ for LHC and $VT^{3}=1020$ for RHIC, as
discussed in section \ref{sec:Results}, together with the value of the second order susceptibility
from  \cite{Borsanyi:2018grb}, $\chi_{2}$, we get the following values for $K_{2}$ and $R_{n,2}$:
\begin{align}
  & \left. K_{2}\right|_{VT^{3}=2560}= 269.3, \,\,\,\,   \left. K_{2} \right|_{VT^{3}=1020} = 107.3, \non
  &R_{4,2}\simeq 0.807, \,\,\,\, R_{6,2} \simeq 0.270, \,\,\,\, R_{8,2}\simeq -0.684 . 
  \label{eq:app:parameters}
\end{align}

Given the above values for $K_{2}$ and $R_{n,2}$ the relative error is dominated by the
leading term, 
\begin{align}
 \frac{\Delta K_{4}}{\left|K_{4}\right| } =\frac{\sqrt{24} }{\sqrt{\neve} }  \frac{K_{2}}{\left|R_{4,2}\right|}.
  \label{eq:k_4_error_leading}
\end{align}
As a consequence, the expected relative error consistent with lattice QCD is roughly 20\% larger
than that of a Skellam distribution of the same width, $\sigma=\sqrt{K_{2}}$, since for the Skellam
distribution $R_{4,2}=1$. Inserting the values for $K_{2}$ and $R_{n,2}$,
Eq.~\eqref{eq:app:parameters}, into Eq.~\eqref{eq:k4_error} we get for the relative error for LHC and RHIC
conditions,
\begin{align}
  \left. \frac{\Delta K_{4}}{\left|K_{4}\right|}\right|_{VT^{3}=2560} &= \frac{1640}{\sqrt{\neve} },\\
  \left. \frac{\Delta K_{4}}{\left|K_{4}\right|}\right|_{VT^{3}=1020} &= \frac{660 }{\sqrt{\neve} },
  \label{eq:k4error_numbers}
\end{align}
so that a measurement with 10\% accuracy would require at least $2.7 \times 10^{8}$  and $4.4 \times 10^{7} $ events for LHC and
RHIC conditions, respectively. 

The expected relative error for the sixth and higher order cumulants are obtained in an analogous fashion and
here we simply give the result for the sixth order cumulant
\begin{align}
  \frac{\Delta K_{6}}{\left|K_{6}\right|}=\frac{\sqrt{720} }{\sqrt{\neve} }  \frac{K^{2}_{2}}{\left|
  R_{6,2}\right| }
   & \left(  1+\frac{15 R_{4,2}}{2 K_2}+\frac{85 R_{4,2}^2}{4 K_2^2}+\frac{55 R_{4,2}^3}{8 K_2^3}+\frac{10 R_{6,2}}{3 K_2^2}+\frac{10 R_{4,2}
  R_{6,2}}{K_2^3} \right.  \non
  & \left. +\frac{461 R_{6,2}^2}{720 K_2^4}+\frac{5 R_{8,2}}{8 K_2^3}+\frac{31 R_{4,2} R_{8,2}}{48 K_2^4}+\frac{R_{10,2}}{20
   K_2^4}+\frac{R_{12,2}}{720 K_2^5} \right)^{1/2} .
  \label{eq:k6_error}
\end{align}
Again we find that the leading term dominates. Also in this case ratios $R_{10,2}$ and $R_{12,2}$
are needed, which are not yet available from lattice QCD. However, as their contributions are
suppressed by factors of $K_{2}^{4}\simeq 10^{8}$ and $K_{2}^{5}\simeq 10^{10}$ it is very unlikely
that they affect the error estimate. Incidentally the two virial expansion models,
Refs. \cite{Vovchenko:2017gkg} and \cite{Almasi:2018lok}  discussed  in this paper  give values of
$R_{10,2} = 0.55$, $R_{12,2}= 18.3$ for model of Ref.~\cite{Vovchenko:2017gkg} and  
$R_{10,2}= 0.45$, $R_{12,2}=1.02$ for the model of Ref.~\cite{Almasi:2018lok}.
Inserting the known values for $R_{n,2}$ and neglecting the contribution from
$R_{10,2}$ and $R_{12,2}$, we get for the relative error:
\begin{align}
  \left. \frac{\Delta K_{6}}{\left|K_{6}\right| }\right|_{VT^{3}=2560} &= \frac{7.3 \times 10^6}{\sqrt{\neve} },\\
  \left. \frac{\Delta K_{6}}{\left|K_{6}\right| }\right|_{VT^{3}=1020} &= \frac{1.2 \times 10^6 }{\sqrt{\neve} } .
  \label{eq:k6error_numbers}
\end{align}
In this case a measurement with 10\% accuracy would require at least 
$5.3 \times 10^{15}$ and $1.4 \times 10^{14}$ for LHC and RHIC conditions, respectively.

Given the above findings, one may also estimate the relative error for net-proton
cumulants, which are more readily accessible in experiment. In this case, the second order cumulant,
$K_{2}^{p}$ has already been determined. Preliminary data from the ALICE
\cite{Rustamov:2017lio} and the STAR \cite{Luo:2015ewa} collaborations give
$K_{2}^{p} \simeq 35$ and $K_{2}^{p} \simeq 25$ for LHC and top 
RHIC energies, respectively. Inserting these values together with the lattice results for
$R_{n,2}$ in Eqs.~(\ref{eq:k4_error}, \ref{eq:k6_error}) we get for the relative errors in
case of the LHC and RHIC
\begin{align}
  \left. \frac{\Delta K_{4}^{p}}{\left| K_{4}^{p} \right|}\right|_{\rm LHC} & = \frac{220}{\sqrt{\neve}}, \,\,\,\,\,\,
  \left. \frac{\Delta K_{6}^{p}}{\left|K_{6}^{p}\right|}\right|_{\rm LHC}  =  \frac{1.3 \times 10^{5}}{\sqrt{\neve}},\\ 
  \left. \frac{\Delta K_{4}^{p}}{\left| K_{4}^{p} \right|}\right|_{\rm RHIC} & = \frac{160}{\sqrt{\neve}}, \,\,\,\,\,\,
  \left. \frac{\Delta K_{6}^{p}}{\left|K_{6}^{p}\right|}\right|_{\rm RHIC}  =  \frac{7 \times 10^{4}}{\sqrt{\neve}}.
 \label{eq:proton_error_numbers}
\end{align}
Thus at LHC a measurement with 10\% accuracy would require $4.8\times 10^{6}$ and $ 1.8\times 10^{12}$ events
for $K_4^p$ and $K_{6}^p$, respectively.
The corresponding values for RHIC  are   $2.5 \times 10^{6}$ and $4.9\times 10^{11}$ events.

However, in case of net-protons, one expects the cumulant ratios to be closer to the Skellam value of $R_{n,2}=1$
since, following the arguments of
\cite{Kitazawa:2011wh}, the net-proton distribution can be derived from the net-baryon distribution by
binomial folding with a Bernoulli probability of $p \simeq 0.5$. In addition, due to $p_{T}$  cuts
not all protons are detected. Both these effects tend to drive the cumulant ratios closer to
$R_{n,2}=1$ \cite{Bzdak:2012ab}. Therefore, a possibly more realistic values for the expected errors are
obtained using $R_{n,2}=1$. In this case we get for the relative errors in case of the LHC:
$\Delta K_{4}^{p}/\left| K_{4}^{p}\right| = 180/\sqrt{\neve}$ and 
$\Delta K_{6}^{p}/\left|K_{6}^{p} \right|= 3.7\times 10^{4}/\sqrt{\neve}$ 
requiring $3.2 \times 10^{6}$ and $1.3 \times 10^{11}$ events for 10\% accuracy, respectively.
The corresponding number of events for RHIC are $1.7\times 10^{6}$ and $ 3.8 \times 10^{10}$.

Finally we note that the leading terms for the relative errors in Eqs.~\eqref{eq:k4_error} and \eqref{eq:k6_error} are within
3\% of the correct value for the net-baryon number cumulants, where $K_{2} \geq 100$ and still
within 12\% for the net-proton cumulants where $K_{2}^{p} \simeq 30$.
Also, this analysis  addresses only the statistical error and thus does not take any 
systematic errors due to, e.g., detection efficiency (see e.g. \cite{Luo:2014rea}).

\bibliography{paper}

\begin{thebibliography}{44}%
\makeatletter
\providecommand \@ifxundefined [1]{%
 \@ifx{#1\undefined}
}%
\providecommand \@ifnum [1]{%
 \ifnum #1\expandafter \@firstoftwo
 \else \expandafter \@secondoftwo
 \fi
}%
\providecommand \@ifx [1]{%
 \ifx #1\expandafter \@firstoftwo
 \else \expandafter \@secondoftwo
 \fi
}%
\providecommand \natexlab [1]{#1}%
\providecommand \enquote  [1]{``#1''}%
\providecommand \bibnamefont  [1]{#1}%
\providecommand \bibfnamefont [1]{#1}%
\providecommand \citenamefont [1]{#1}%
\providecommand \href@noop [0]{\@secondoftwo}%
\providecommand \href [0]{\begingroup \@sanitize@url \@href}%
\providecommand \@href[1]{\@@startlink{#1}\@@href}%
\providecommand \@@href[1]{\endgroup#1\@@endlink}%
\providecommand \@sanitize@url [0]{\catcode `\\12\catcode `\$12\catcode
  `\&12\catcode `\#12\catcode `\^12\catcode `\_12\catcode `\%12\relax}%
\providecommand \@@startlink[1]{}%
\providecommand \@@endlink[0]{}%
\providecommand \url  [0]{\begingroup\@sanitize@url \@url }%
\providecommand \@url [1]{\endgroup\@href {#1}{\urlprefix }}%
\providecommand \urlprefix  [0]{URL }%
\providecommand \Eprint [0]{\href }%
\providecommand \doibase [0]{http://dx.doi.org/}%
\providecommand \selectlanguage [0]{\@gobble}%
\providecommand \bibinfo  [0]{\@secondoftwo}%
\providecommand \bibfield  [0]{\@secondoftwo}%
\providecommand \translation [1]{[#1]}%
\providecommand \BibitemOpen [0]{}%
\providecommand \bibitemStop [0]{}%
\providecommand \bibitemNoStop [0]{.\EOS\space}%
\providecommand \EOS [0]{\spacefactor3000\relax}%
\providecommand \BibitemShut  [1]{\csname bibitem#1\endcsname}%
\let\auto@bib@innerbib\@empty
\bibitem [{\citenamefont {Jeon}\ and\ \citenamefont
  {Koch}(2000)}]{Jeon:2000wg}%
  \BibitemOpen
  \bibfield  {author} {\bibinfo {author} {\bibfnamefont {S.}~\bibnamefont
  {Jeon}}\ and\ \bibinfo {author} {\bibfnamefont {V.}~\bibnamefont {Koch}},\
  }\href {\doibase 10.1103/PhysRevLett.85.2076} {\bibfield  {journal} {\bibinfo
   {journal} {Phys. Rev. Lett.}\ }\textbf {\bibinfo {volume} {85}},\ \bibinfo
  {pages} {2076} (\bibinfo {year} {2000})},\ \Eprint
  {http://arxiv.org/abs/hep-ph/0003168} {arXiv:hep-ph/0003168 [hep-ph]}
  \BibitemShut {NoStop}%
\bibitem [{\citenamefont {Asakawa}\ \emph {et~al.}(2000)\citenamefont
  {Asakawa}, \citenamefont {Heinz},\ and\ \citenamefont
  {Muller}}]{Asakawa:2000wh}%
  \BibitemOpen
  \bibfield  {author} {\bibinfo {author} {\bibfnamefont {M.}~\bibnamefont
  {Asakawa}}, \bibinfo {author} {\bibfnamefont {U.~W.}\ \bibnamefont {Heinz}},
  \ and\ \bibinfo {author} {\bibfnamefont {B.}~\bibnamefont {Muller}},\
  }\href@noop {} {\bibfield  {journal} {\bibinfo  {journal} {Phys. Rev. Lett.}\
  }\textbf {\bibinfo {volume} {85}},\ \bibinfo {pages} {2072} (\bibinfo {year}
  {2000})},\ \Eprint {http://arXiv.org/abs/hep-ph/0003169} {hep-ph/0003169}
  \BibitemShut {NoStop}%
\bibitem [{\citenamefont {Stephanov}(2009)}]{Stephanov:2008qz}%
  \BibitemOpen
  \bibfield  {author} {\bibinfo {author} {\bibfnamefont {M.}~\bibnamefont
  {Stephanov}},\ }\href {\doibase 10.1103/PhysRevLett.102.032301} {\bibfield
  {journal} {\bibinfo  {journal} {Phys.Rev.Lett.}\ }\textbf {\bibinfo {volume}
  {102}},\ \bibinfo {pages} {032301} (\bibinfo {year} {2009})},\ \Eprint
  {http://arxiv.org/abs/0809.3450} {arXiv:0809.3450 [hep-ph]} \BibitemShut
  {NoStop}%
\bibitem [{\citenamefont {Skokov}\ \emph {et~al.}(2011)\citenamefont {Skokov},
  \citenamefont {Friman},\ and\ \citenamefont {Redlich}}]{Skokov:2010uh}%
  \BibitemOpen
  \bibfield  {author} {\bibinfo {author} {\bibfnamefont {V.}~\bibnamefont
  {Skokov}}, \bibinfo {author} {\bibfnamefont {B.}~\bibnamefont {Friman}}, \
  and\ \bibinfo {author} {\bibfnamefont {K.}~\bibnamefont {Redlich}},\ }\href
  {\doibase 10.1103/PhysRevC.83.054904} {\bibfield  {journal} {\bibinfo
  {journal} {Phys.Rev.}\ }\textbf {\bibinfo {volume} {C83}},\ \bibinfo {pages}
  {054904} (\bibinfo {year} {2011})},\ \Eprint {http://arxiv.org/abs/1008.4570}
  {arXiv:1008.4570 [hep-ph]} \BibitemShut {NoStop}%
\bibitem [{\citenamefont {Stephanov}(2011)}]{Stephanov:2011pb}%
  \BibitemOpen
  \bibfield  {author} {\bibinfo {author} {\bibfnamefont {M.}~\bibnamefont
  {Stephanov}},\ }\href {\doibase 10.1103/PhysRevLett.107.052301} {\bibfield
  {journal} {\bibinfo  {journal} {Phys.Rev.Lett.}\ }\textbf {\bibinfo {volume}
  {107}},\ \bibinfo {pages} {052301} (\bibinfo {year} {2011})},\ \Eprint
  {http://arxiv.org/abs/1104.1627} {arXiv:1104.1627 [hep-ph]} \BibitemShut
  {NoStop}%
\bibitem [{\citenamefont {Luo}\ \emph {et~al.}(2012)\citenamefont {Luo},
  \citenamefont {Mohanty}, \citenamefont {Ritter},\ and\ \citenamefont
  {Xu}}]{Luo:2011rg}%
  \BibitemOpen
  \bibfield  {author} {\bibinfo {author} {\bibfnamefont {X.-F.}\ \bibnamefont
  {Luo}}, \bibinfo {author} {\bibfnamefont {B.}~\bibnamefont {Mohanty}},
  \bibinfo {author} {\bibfnamefont {H.~G.}\ \bibnamefont {Ritter}}, \ and\
  \bibinfo {author} {\bibfnamefont {N.}~\bibnamefont {Xu}},\ }\href {\doibase
  10.1134/S1063778812060348} {\bibfield  {journal} {\bibinfo  {journal} {Phys.
  Atom. Nucl.}\ }\textbf {\bibinfo {volume} {75}},\ \bibinfo {pages} {676}
  (\bibinfo {year} {2012})},\ \Eprint {http://arxiv.org/abs/1105.5049}
  {arXiv:1105.5049 [nucl-ex]} \BibitemShut {NoStop}%
\bibitem [{\citenamefont {Luo}\ and\ \citenamefont {Xu}(2017)}]{Luo:2017faz}%
  \BibitemOpen
  \bibfield  {author} {\bibinfo {author} {\bibfnamefont {X.}~\bibnamefont
  {Luo}}\ and\ \bibinfo {author} {\bibfnamefont {N.}~\bibnamefont {Xu}},\
  }\href {\doibase 10.1007/s41365-017-0257-0} {\bibfield  {journal} {\bibinfo
  {journal} {Nucl. Sci. Tech.}\ }\textbf {\bibinfo {volume} {28}},\ \bibinfo
  {pages} {112} (\bibinfo {year} {2017})},\ \Eprint
  {http://arxiv.org/abs/1701.02105} {arXiv:1701.02105 [nucl-ex]} \BibitemShut
  {NoStop}%
\bibitem [{\citenamefont {Herold}\ \emph {et~al.}(2016)\citenamefont {Herold},
  \citenamefont {Nahrgang}, \citenamefont {Yan},\ and\ \citenamefont
  {Kobdaj}}]{Herold:2016uvv}%
  \BibitemOpen
  \bibfield  {author} {\bibinfo {author} {\bibfnamefont {C.}~\bibnamefont
  {Herold}}, \bibinfo {author} {\bibfnamefont {M.}~\bibnamefont {Nahrgang}},
  \bibinfo {author} {\bibfnamefont {Y.}~\bibnamefont {Yan}}, \ and\ \bibinfo
  {author} {\bibfnamefont {C.}~\bibnamefont {Kobdaj}},\ }\href {\doibase
  10.1103/PhysRevC.93.021902} {\bibfield  {journal} {\bibinfo  {journal} {Phys.
  Rev.}\ }\textbf {\bibinfo {volume} {C93}},\ \bibinfo {pages} {021902}
  (\bibinfo {year} {2016})},\ \Eprint {http://arxiv.org/abs/1601.04839}
  {arXiv:1601.04839 [hep-ph]} \BibitemShut {NoStop}%
\bibitem [{\citenamefont {Zhou}\ \emph {et~al.}(2012)\citenamefont {Zhou},
  \citenamefont {Limphirat}, \citenamefont {Yan}, \citenamefont {Yun},
  \citenamefont {Yan}, \citenamefont {Cai}, \citenamefont {Csernai},\ and\
  \citenamefont {Sa}}]{Zhou:2012ay}%
  \BibitemOpen
  \bibfield  {author} {\bibinfo {author} {\bibfnamefont {D.-M.}\ \bibnamefont
  {Zhou}}, \bibinfo {author} {\bibfnamefont {A.}~\bibnamefont {Limphirat}},
  \bibinfo {author} {\bibfnamefont {Y.-l.}\ \bibnamefont {Yan}}, \bibinfo
  {author} {\bibfnamefont {C.}~\bibnamefont {Yun}}, \bibinfo {author}
  {\bibfnamefont {Y.-p.}\ \bibnamefont {Yan}}, \bibinfo {author} {\bibfnamefont
  {X.}~\bibnamefont {Cai}}, \bibinfo {author} {\bibfnamefont {L.~P.}\
  \bibnamefont {Csernai}}, \ and\ \bibinfo {author} {\bibfnamefont {B.-H.}\
  \bibnamefont {Sa}},\ }\href {\doibase 10.1103/PhysRevC.85.064916} {\bibfield
  {journal} {\bibinfo  {journal} {Phys. Rev.}\ }\textbf {\bibinfo {volume}
  {C85}},\ \bibinfo {pages} {064916} (\bibinfo {year} {2012})},\ \Eprint
  {http://arxiv.org/abs/1205.5634} {arXiv:1205.5634 [nucl-th]} \BibitemShut
  {NoStop}%
\bibitem [{\citenamefont {Wang}\ and\ \citenamefont
  {Yang}(2012)}]{Wang:2012jr}%
  \BibitemOpen
  \bibfield  {author} {\bibinfo {author} {\bibfnamefont {X.}~\bibnamefont
  {Wang}}\ and\ \bibinfo {author} {\bibfnamefont {C.~B.}\ \bibnamefont
  {Yang}},\ }\href {\doibase 10.1103/PhysRevC.85.044905} {\bibfield  {journal}
  {\bibinfo  {journal} {Phys. Rev.}\ }\textbf {\bibinfo {volume} {C85}},\
  \bibinfo {pages} {044905} (\bibinfo {year} {2012})},\ \Eprint
  {http://arxiv.org/abs/1202.4857} {arXiv:1202.4857 [nucl-th]} \BibitemShut
  {NoStop}%
\bibitem [{\citenamefont {Karsch}\ and\ \citenamefont
  {Redlich}(2011)}]{Karsch:2011gg}%
  \BibitemOpen
  \bibfield  {author} {\bibinfo {author} {\bibfnamefont {F.}~\bibnamefont
  {Karsch}}\ and\ \bibinfo {author} {\bibfnamefont {K.}~\bibnamefont
  {Redlich}},\ }\href {\doibase 10.1103/PhysRevD.84.051504} {\bibfield
  {journal} {\bibinfo  {journal} {Phys. Rev.}\ }\textbf {\bibinfo {volume}
  {D84}},\ \bibinfo {pages} {051504} (\bibinfo {year} {2011})},\ \Eprint
  {http://arxiv.org/abs/1107.1412} {arXiv:1107.1412 [hep-ph]} \BibitemShut
  {NoStop}%
\bibitem [{\citenamefont {Schaefer}\ and\ \citenamefont
  {Wagner}(2012)}]{Schaefer:2011ex}%
  \BibitemOpen
  \bibfield  {author} {\bibinfo {author} {\bibfnamefont {B.~J.}\ \bibnamefont
  {Schaefer}}\ and\ \bibinfo {author} {\bibfnamefont {M.}~\bibnamefont
  {Wagner}},\ }\href {\doibase 10.1103/PhysRevD.85.034027} {\bibfield
  {journal} {\bibinfo  {journal} {Phys. Rev.}\ }\textbf {\bibinfo {volume}
  {D85}},\ \bibinfo {pages} {034027} (\bibinfo {year} {2012})},\ \Eprint
  {http://arxiv.org/abs/1111.6871} {arXiv:1111.6871 [hep-ph]} \BibitemShut
  {NoStop}%
\bibitem [{\citenamefont {Chen}\ \emph {et~al.}(2011)\citenamefont {Chen},
  \citenamefont {Pan}, \citenamefont {Xiong}, \citenamefont {Li}, \citenamefont
  {Li}, \citenamefont {Li}, \citenamefont {Wang},\ and\ \citenamefont
  {Wu}}]{Chen:2011am}%
  \BibitemOpen
  \bibfield  {author} {\bibinfo {author} {\bibfnamefont {L.}~\bibnamefont
  {Chen}}, \bibinfo {author} {\bibfnamefont {X.}~\bibnamefont {Pan}}, \bibinfo
  {author} {\bibfnamefont {F.-B.}\ \bibnamefont {Xiong}}, \bibinfo {author}
  {\bibfnamefont {L.}~\bibnamefont {Li}}, \bibinfo {author} {\bibfnamefont
  {N.}~\bibnamefont {Li}}, \bibinfo {author} {\bibfnamefont {Z.}~\bibnamefont
  {Li}}, \bibinfo {author} {\bibfnamefont {G.}~\bibnamefont {Wang}}, \ and\
  \bibinfo {author} {\bibfnamefont {Y.}~\bibnamefont {Wu}},\ }\href {\doibase
  10.1088/0954-3899/38/11/115004} {\bibfield  {journal} {\bibinfo  {journal}
  {J. Phys.}\ }\textbf {\bibinfo {volume} {G38}},\ \bibinfo {pages} {115004}
  (\bibinfo {year} {2011})}\BibitemShut {NoStop}%
\bibitem [{\citenamefont {Fu}\ \emph {et~al.}(2010)\citenamefont {Fu},
  \citenamefont {Liu},\ and\ \citenamefont {Wu}}]{Fu:2009wy}%
  \BibitemOpen
  \bibfield  {author} {\bibinfo {author} {\bibfnamefont {W.-j.}\ \bibnamefont
  {Fu}}, \bibinfo {author} {\bibfnamefont {Y.-x.}\ \bibnamefont {Liu}}, \ and\
  \bibinfo {author} {\bibfnamefont {Y.-L.}\ \bibnamefont {Wu}},\ }\href
  {\doibase 10.1103/PhysRevD.81.014028} {\bibfield  {journal} {\bibinfo
  {journal} {Phys. Rev.}\ }\textbf {\bibinfo {volume} {D81}},\ \bibinfo {pages}
  {014028} (\bibinfo {year} {2010})},\ \Eprint {http://arxiv.org/abs/0910.5783}
  {arXiv:0910.5783 [hep-ph]} \BibitemShut {NoStop}%
\bibitem [{\citenamefont {Cheng}\ \emph {et~al.}(2009)\citenamefont {Cheng}
  \emph {et~al.}}]{Cheng:2008zh}%
  \BibitemOpen
  \bibfield  {author} {\bibinfo {author} {\bibfnamefont {M.}~\bibnamefont
  {Cheng}} \emph {et~al.},\ }\href@noop {} {\bibfield  {journal} {\bibinfo
  {journal} {Phys. Rev.}\ }\textbf {\bibinfo {volume} {D79}},\ \bibinfo {pages}
  {074505} (\bibinfo {year} {2009})},\ \Eprint {http://arxiv.org/abs/0811.1006}
  {arXiv:0811.1006 [hep-lat]} \BibitemShut {NoStop}%
\bibitem [{\citenamefont {Aoki}\ \emph {et~al.}(2006)\citenamefont {Aoki},
  \citenamefont {Endrodi}, \citenamefont {Fodor}, \citenamefont {Katz},\ and\
  \citenamefont {Szabo}}]{Aoki:2006we}%
  \BibitemOpen
  \bibfield  {author} {\bibinfo {author} {\bibfnamefont {Y.}~\bibnamefont
  {Aoki}}, \bibinfo {author} {\bibfnamefont {G.}~\bibnamefont {Endrodi}},
  \bibinfo {author} {\bibfnamefont {Z.}~\bibnamefont {Fodor}}, \bibinfo
  {author} {\bibfnamefont {S.~D.}\ \bibnamefont {Katz}}, \ and\ \bibinfo
  {author} {\bibfnamefont {K.~K.}\ \bibnamefont {Szabo}},\ }\href {\doibase
  10.1038/nature05120} {\bibfield  {journal} {\bibinfo  {journal} {Nature}\
  }\textbf {\bibinfo {volume} {443}},\ \bibinfo {pages} {675} (\bibinfo {year}
  {2006})},\ \Eprint {http://arxiv.org/abs/hep-lat/0611014}
  {arXiv:hep-lat/0611014} \BibitemShut {NoStop}%
\bibitem [{\citenamefont {Borsanyi}\ \emph {et~al.}(2010)\citenamefont
  {Borsanyi}, \citenamefont {Endrodi}, \citenamefont {Fodor}, \citenamefont
  {Jakovac}, \citenamefont {Katz} \emph {et~al.}}]{Borsanyi:2010cj}%
  \BibitemOpen
  \bibfield  {author} {\bibinfo {author} {\bibfnamefont {S.}~\bibnamefont
  {Borsanyi}}, \bibinfo {author} {\bibfnamefont {G.}~\bibnamefont {Endrodi}},
  \bibinfo {author} {\bibfnamefont {Z.}~\bibnamefont {Fodor}}, \bibinfo
  {author} {\bibfnamefont {A.}~\bibnamefont {Jakovac}}, \bibinfo {author}
  {\bibfnamefont {S.~D.}\ \bibnamefont {Katz}},  \emph {et~al.},\ }\href
  {\doibase 10.1007/JHEP11(2010)077} {\bibfield  {journal} {\bibinfo  {journal}
  {JHEP}\ }\textbf {\bibinfo {volume} {1011}},\ \bibinfo {pages} {077}
  (\bibinfo {year} {2010})},\ \Eprint {http://arxiv.org/abs/1007.2580}
  {arXiv:1007.2580 [hep-lat]} \BibitemShut {NoStop}%
\bibitem [{\citenamefont {Bazavov}\ \emph {et~al.}(2012)\citenamefont
  {Bazavov}, \citenamefont {Bhattacharya}, \citenamefont {Cheng}, \citenamefont
  {DeTar}, \citenamefont {Ding}, \citenamefont {Karsch} \emph
  {et~al.}}]{Bazavov:2011nk}%
  \BibitemOpen
  \bibfield  {author} {\bibinfo {author} {\bibfnamefont {A.}~\bibnamefont
  {Bazavov}}, \bibinfo {author} {\bibfnamefont {T.}~\bibnamefont
  {Bhattacharya}}, \bibinfo {author} {\bibfnamefont {M.}~\bibnamefont {Cheng}},
  \bibinfo {author} {\bibfnamefont {C.}~\bibnamefont {DeTar}}, \bibinfo
  {author} {\bibfnamefont {H.}~\bibnamefont {Ding}}, \bibinfo {author}
  {\bibfnamefont {F.}~\bibnamefont {Karsch}},  \emph {et~al.},\ }\href
  {\doibase 10.1103/PhysRevD.85.054503} {\bibfield  {journal} {\bibinfo
  {journal} {Phys.Rev.}\ }\textbf {\bibinfo {volume} {D85}},\ \bibinfo {pages}
  {054503} (\bibinfo {year} {2012})},\ \Eprint {http://arxiv.org/abs/1111.1710}
  {arXiv:1111.1710 [hep-lat]} \BibitemShut {NoStop}%
\bibitem [{\citenamefont {Adamczyk}\ \emph {et~al.}(2014)\citenamefont
  {Adamczyk} \emph {et~al.}}]{Adamczyk:2013dal}%
  \BibitemOpen
  \bibfield  {author} {\bibinfo {author} {\bibfnamefont {L.}~\bibnamefont
  {Adamczyk}} \emph {et~al.} (\bibinfo {collaboration} {STAR}),\ }\href
  {\doibase 10.1103/PhysRevLett.112.032302} {\bibfield  {journal} {\bibinfo
  {journal} {Phys. Rev. Lett.}\ }\textbf {\bibinfo {volume} {112}},\ \bibinfo
  {pages} {032302} (\bibinfo {year} {2014})},\ \Eprint
  {http://arxiv.org/abs/1309.5681} {arXiv:1309.5681 [nucl-ex]} \BibitemShut
  {NoStop}%
\bibitem [{\citenamefont {Rustamov}(2017)}]{Rustamov:2017lio}%
  \BibitemOpen
  \bibfield  {author} {\bibinfo {author} {\bibfnamefont {A.}~\bibnamefont
  {Rustamov}} (\bibinfo {collaboration} {ALICE}),\ }\bibfield  {booktitle}
  {\emph {\bibinfo {booktitle} {{Proceedings, 26th International Conference on
  Ultra-relativistic Nucleus-Nucleus Collisions (Quark Matter 2017): Chicago,
  Illinois, USA, February 5-11, 2017}}},\ }\href {\doibase
  10.1016/j.nuclphysa.2017.05.111} {\bibfield  {journal} {\bibinfo  {journal}
  {Nucl. Phys.}\ }\textbf {\bibinfo {volume} {A967}},\ \bibinfo {pages} {453}
  (\bibinfo {year} {2017})},\ \Eprint {http://arxiv.org/abs/1704.05329}
  {arXiv:1704.05329 [nucl-ex]} \BibitemShut {NoStop}%
\bibitem [{\citenamefont {Kitazawa}\ and\ \citenamefont
  {Asakawa}(2012{\natexlab{a}})}]{Kitazawa:2011wh}%
  \BibitemOpen
  \bibfield  {author} {\bibinfo {author} {\bibfnamefont {M.}~\bibnamefont
  {Kitazawa}}\ and\ \bibinfo {author} {\bibfnamefont {M.}~\bibnamefont
  {Asakawa}},\ }\href {\doibase 10.1103/PhysRevC.85.021901} {\bibfield
  {journal} {\bibinfo  {journal} {Phys. Rev.}\ }\textbf {\bibinfo {volume}
  {C85}},\ \bibinfo {pages} {021901} (\bibinfo {year} {2012}{\natexlab{a}})},\
  \Eprint {http://arxiv.org/abs/1107.2755} {arXiv:1107.2755 [nucl-th]}
  \BibitemShut {NoStop}%
\bibitem [{\citenamefont {Kitazawa}\ and\ \citenamefont
  {Asakawa}(2012{\natexlab{b}})}]{Kitazawa:2012at}%
  \BibitemOpen
  \bibfield  {author} {\bibinfo {author} {\bibfnamefont {M.}~\bibnamefont
  {Kitazawa}}\ and\ \bibinfo {author} {\bibfnamefont {M.}~\bibnamefont
  {Asakawa}},\ }\href {\doibase 10.1103/PhysRevC.86.024904,
  10.1103/PhysRevC.86.069902} {\bibfield  {journal} {\bibinfo  {journal}
  {Phys.Rev.}\ }\textbf {\bibinfo {volume} {C86}},\ \bibinfo {pages} {024904}
  (\bibinfo {year} {2012}{\natexlab{b}})},\ \Eprint
  {http://arxiv.org/abs/1205.3292} {arXiv:1205.3292 [nucl-th]} \BibitemShut
  {NoStop}%
\bibitem [{\citenamefont {Bzdak}\ \emph {et~al.}(2013)\citenamefont {Bzdak},
  \citenamefont {Koch},\ and\ \citenamefont {Skokov}}]{Bzdak:2012an}%
  \BibitemOpen
  \bibfield  {author} {\bibinfo {author} {\bibfnamefont {A.}~\bibnamefont
  {Bzdak}}, \bibinfo {author} {\bibfnamefont {V.}~\bibnamefont {Koch}}, \ and\
  \bibinfo {author} {\bibfnamefont {V.}~\bibnamefont {Skokov}},\ }\href
  {\doibase 10.1103/PhysRevC.87.014901} {\bibfield  {journal} {\bibinfo
  {journal} {Phys. Rev.}\ }\textbf {\bibinfo {volume} {C87}},\ \bibinfo {pages}
  {014901} (\bibinfo {year} {2013})},\ \Eprint {http://arxiv.org/abs/1203.4529}
  {arXiv:1203.4529 [hep-ph]} \BibitemShut {NoStop}%
\bibitem [{\citenamefont {Bzdak}\ and\ \citenamefont
  {Koch}(2012)}]{Bzdak:2012ab}%
  \BibitemOpen
  \bibfield  {author} {\bibinfo {author} {\bibfnamefont {A.}~\bibnamefont
  {Bzdak}}\ and\ \bibinfo {author} {\bibfnamefont {V.}~\bibnamefont {Koch}},\
  }\href {\doibase 10.1103/PhysRevC.86.044904} {\bibfield  {journal} {\bibinfo
  {journal} {Phys. Rev.}\ }\textbf {\bibinfo {volume} {C86}},\ \bibinfo {pages}
  {044904} (\bibinfo {year} {2012})},\ \Eprint {http://arxiv.org/abs/1206.4286}
  {arXiv:1206.4286 [nucl-th]} \BibitemShut {NoStop}%
\bibitem [{\citenamefont {Luo}(2015{\natexlab{a}})}]{Luo:2014rea}%
  \BibitemOpen
  \bibfield  {author} {\bibinfo {author} {\bibfnamefont {X.}~\bibnamefont
  {Luo}},\ }\href {\doibase 10.1103/PhysRevC.91.034907} {\bibfield  {journal}
  {\bibinfo  {journal} {Phys. Rev.}\ }\textbf {\bibinfo {volume} {C91}},\
  \bibinfo {pages} {034907} (\bibinfo {year} {2015}{\natexlab{a}})},\ \Eprint
  {http://arxiv.org/abs/1410.3914} {arXiv:1410.3914 [physics.data-an]}
  \BibitemShut {NoStop}%
\bibitem [{\citenamefont {Bzdak}\ \emph {et~al.}(2016)\citenamefont {Bzdak},
  \citenamefont {Holzmann},\ and\ \citenamefont {Koch}}]{Bzdak:2016qdc}%
  \BibitemOpen
  \bibfield  {author} {\bibinfo {author} {\bibfnamefont {A.}~\bibnamefont
  {Bzdak}}, \bibinfo {author} {\bibfnamefont {R.}~\bibnamefont {Holzmann}}, \
  and\ \bibinfo {author} {\bibfnamefont {V.}~\bibnamefont {Koch}},\ }\href
  {\doibase 10.1103/PhysRevC.94.064907} {\bibfield  {journal} {\bibinfo
  {journal} {Phys. Rev.}\ }\textbf {\bibinfo {volume} {C94}},\ \bibinfo {pages}
  {064907} (\bibinfo {year} {2016})},\ \Eprint
  {http://arxiv.org/abs/1603.09057} {arXiv:1603.09057 [nucl-th]} \BibitemShut
  {NoStop}%
\bibitem [{\citenamefont {Braun-Munzinger}\ \emph {et~al.}(2016)\citenamefont
  {Braun-Munzinger}, \citenamefont {Koch}, \citenamefont {Sch{\"a}fer},\ and\
  \citenamefont {Stachel}}]{Braun-Munzinger:2015hba}%
  \BibitemOpen
  \bibfield  {author} {\bibinfo {author} {\bibfnamefont {P.}~\bibnamefont
  {Braun-Munzinger}}, \bibinfo {author} {\bibfnamefont {V.}~\bibnamefont
  {Koch}}, \bibinfo {author} {\bibfnamefont {T.}~\bibnamefont {Sch{\"a}fer}}, \
  and\ \bibinfo {author} {\bibfnamefont {J.}~\bibnamefont {Stachel}},\ }\href
  {\doibase 10.1016/j.physrep.2015.12.003} {\bibfield  {journal} {\bibinfo
  {journal} {Phys. Rept.}\ }\textbf {\bibinfo {volume} {621}},\ \bibinfo
  {pages} {76} (\bibinfo {year} {2016})},\ \Eprint
  {http://arxiv.org/abs/1510.00442} {arXiv:1510.00442 [nucl-th]} \BibitemShut
  {NoStop}%
\bibitem [{\citenamefont {Braun-Munzinger}\ \emph {et~al.}(2017)\citenamefont
  {Braun-Munzinger}, \citenamefont {Rustamov},\ and\ \citenamefont
  {Stachel}}]{Braun-Munzinger:2016yjz}%
  \BibitemOpen
  \bibfield  {author} {\bibinfo {author} {\bibfnamefont {P.}~\bibnamefont
  {Braun-Munzinger}}, \bibinfo {author} {\bibfnamefont {A.}~\bibnamefont
  {Rustamov}}, \ and\ \bibinfo {author} {\bibfnamefont {J.}~\bibnamefont
  {Stachel}},\ }\href {\doibase 10.1016/j.nuclphysa.2017.01.011} {\bibfield
  {journal} {\bibinfo  {journal} {Nucl. Phys.}\ }\textbf {\bibinfo {volume}
  {A960}},\ \bibinfo {pages} {114} (\bibinfo {year} {2017})},\ \Eprint
  {http://arxiv.org/abs/1612.00702} {arXiv:1612.00702 [nucl-th]} \BibitemShut
  {NoStop}%
\bibitem [{\citenamefont {Kitazawa}(2016)}]{Kitazawa:2016awu}%
  \BibitemOpen
  \bibfield  {author} {\bibinfo {author} {\bibfnamefont {M.}~\bibnamefont
  {Kitazawa}},\ }\href {\doibase 10.1103/PhysRevC.93.044911} {\bibfield
  {journal} {\bibinfo  {journal} {Phys. Rev.}\ }\textbf {\bibinfo {volume}
  {C93}},\ \bibinfo {pages} {044911} (\bibinfo {year} {2016})},\ \Eprint
  {http://arxiv.org/abs/1602.01234} {arXiv:1602.01234 [nucl-th]} \BibitemShut
  {NoStop}%
\bibitem [{\citenamefont {He}\ and\ \citenamefont {Luo}(2018)}]{He:2018mri}%
  \BibitemOpen
  \bibfield  {author} {\bibinfo {author} {\bibfnamefont {S.}~\bibnamefont
  {He}}\ and\ \bibinfo {author} {\bibfnamefont {X.}~\bibnamefont {Luo}},\
  }\href {\doibase 10.1088/1674-1137/42/10/104001} {\bibfield  {journal}
  {\bibinfo  {journal} {Chin. Phys.}\ }\textbf {\bibinfo {volume} {C42}},\
  \bibinfo {pages} {104001} (\bibinfo {year} {2018})},\ \Eprint
  {http://arxiv.org/abs/1802.02911} {arXiv:1802.02911 [physics.data-an]}
  \BibitemShut {NoStop}%
\bibitem [{\citenamefont {Nonaka}\ \emph {et~al.}(2018)\citenamefont {Nonaka},
  \citenamefont {Kitazawa},\ and\ \citenamefont {Esumi}}]{Nonaka:2018mgw}%
  \BibitemOpen
  \bibfield  {author} {\bibinfo {author} {\bibfnamefont {T.}~\bibnamefont
  {Nonaka}}, \bibinfo {author} {\bibfnamefont {M.}~\bibnamefont {Kitazawa}}, \
  and\ \bibinfo {author} {\bibfnamefont {S.}~\bibnamefont {Esumi}},\ }\href
  {\doibase 10.1016/j.nima.2018.08.013} {\bibfield  {journal} {\bibinfo
  {journal} {Nucl. Instrum. Meth.}\ }\textbf {\bibinfo {volume} {A906}},\
  \bibinfo {pages} {10} (\bibinfo {year} {2018})},\ \Eprint
  {http://arxiv.org/abs/1805.00279} {arXiv:1805.00279 [physics.data-an]}
  \BibitemShut {NoStop}%
\bibitem [{\citenamefont {Bazavov}\ \emph {et~al.}(2017)\citenamefont {Bazavov}
  \emph {et~al.}}]{Bazavov:2017dus}%
  \BibitemOpen
  \bibfield  {author} {\bibinfo {author} {\bibfnamefont {A.}~\bibnamefont
  {Bazavov}} \emph {et~al.},\ }\href {\doibase 10.1103/PhysRevD.95.054504}
  {\bibfield  {journal} {\bibinfo  {journal} {Phys. Rev.}\ }\textbf {\bibinfo
  {volume} {D95}},\ \bibinfo {pages} {054504} (\bibinfo {year} {2017})},\
  \Eprint {http://arxiv.org/abs/1701.04325} {arXiv:1701.04325 [hep-lat]}
  \BibitemShut {NoStop}%
\bibitem [{\citenamefont {Borsanyi}\ \emph {et~al.}(2018)\citenamefont
  {Borsanyi}, \citenamefont {Fodor}, \citenamefont {Guenther}, \citenamefont
  {Katz}, \citenamefont {Szab{\'o}}, \citenamefont {Pasztor}, \citenamefont
  {Portillo},\ and\ \citenamefont {Ratti}}]{Borsanyi:2018grb}%
  \BibitemOpen
  \bibfield  {author} {\bibinfo {author} {\bibfnamefont {S.}~\bibnamefont
  {Borsanyi}}, \bibinfo {author} {\bibfnamefont {Z.}~\bibnamefont {Fodor}},
  \bibinfo {author} {\bibfnamefont {J.~N.}\ \bibnamefont {Guenther}}, \bibinfo
  {author} {\bibfnamefont {S.~K.}\ \bibnamefont {Katz}}, \bibinfo {author}
  {\bibfnamefont {K.~K.}\ \bibnamefont {Szab{\'o}}}, \bibinfo {author}
  {\bibfnamefont {A.}~\bibnamefont {Pasztor}}, \bibinfo {author} {\bibfnamefont
  {I.}~\bibnamefont {Portillo}}, \ and\ \bibinfo {author} {\bibfnamefont
  {C.}~\bibnamefont {Ratti}},\ }\href@noop {} {\  (\bibinfo {year} {2018})},\
  \Eprint {http://arxiv.org/abs/1805.04445} {arXiv:1805.04445 [hep-lat]}
  \BibitemShut {NoStop}%
\bibitem [{\citenamefont {Davison}(2003)}]{davison2003statistical}%
  \BibitemOpen
  \bibfield  {author} {\bibinfo {author} {\bibfnamefont {A.}~\bibnamefont
  {Davison}},\ }\href {https://books.google.com/books?id=gQyIGGAiN4AC} {\emph
  {\bibinfo {title} {Statistical Models}}},\ Cambridge Series in Statistical
  and Probabilistic Mathematics\ (\bibinfo  {publisher} {Cambridge University
  Press},\ \bibinfo {year} {2003})\BibitemShut {NoStop}%
\bibitem [{\citenamefont {Morita}\ \emph {et~al.}(2013)\citenamefont {Morita},
  \citenamefont {Friman}, \citenamefont {Redlich},\ and\ \citenamefont
  {Skokov}}]{Morita:2013tu}%
  \BibitemOpen
  \bibfield  {author} {\bibinfo {author} {\bibfnamefont {K.}~\bibnamefont
  {Morita}}, \bibinfo {author} {\bibfnamefont {B.}~\bibnamefont {Friman}},
  \bibinfo {author} {\bibfnamefont {K.}~\bibnamefont {Redlich}}, \ and\
  \bibinfo {author} {\bibfnamefont {V.}~\bibnamefont {Skokov}},\ }\href
  {\doibase 10.1103/PhysRevC.88.034903} {\bibfield  {journal} {\bibinfo
  {journal} {Phys. Rev.}\ }\textbf {\bibinfo {volume} {C88}},\ \bibinfo {pages}
  {034903} (\bibinfo {year} {2013})},\ \Eprint {http://arxiv.org/abs/1301.2873}
  {arXiv:1301.2873 [hep-ph]} \BibitemShut {NoStop}%
\bibitem [{\citenamefont {Braun-Munzinger}\ \emph {et~al.}(2011)\citenamefont
  {Braun-Munzinger}, \citenamefont {Friman}, \citenamefont {Karsch},
  \citenamefont {Redlich},\ and\ \citenamefont
  {Skokov}}]{BraunMunzinger:2011dn}%
  \BibitemOpen
  \bibfield  {author} {\bibinfo {author} {\bibfnamefont {P.}~\bibnamefont
  {Braun-Munzinger}}, \bibinfo {author} {\bibfnamefont {B.}~\bibnamefont
  {Friman}}, \bibinfo {author} {\bibfnamefont {F.}~\bibnamefont {Karsch}},
  \bibinfo {author} {\bibfnamefont {K.}~\bibnamefont {Redlich}}, \ and\
  \bibinfo {author} {\bibfnamefont {V.}~\bibnamefont {Skokov}},\ }\href
  {\doibase 10.1103/PhysRevC.84.064911} {\bibfield  {journal} {\bibinfo
  {journal} {Phys.Rev.}\ }\textbf {\bibinfo {volume} {C84}},\ \bibinfo {pages}
  {064911} (\bibinfo {year} {2011})},\ \Eprint {http://arxiv.org/abs/1107.4267}
  {arXiv:1107.4267 [hep-ph]} \BibitemShut {NoStop}%
\bibitem [{\citenamefont {Vovchenko}\ \emph {et~al.}(2017)\citenamefont
  {Vovchenko}, \citenamefont {Pasztor}, \citenamefont {Fodor}, \citenamefont
  {Katz},\ and\ \citenamefont {Stoecker}}]{Vovchenko:2017xad}%
  \BibitemOpen
  \bibfield  {author} {\bibinfo {author} {\bibfnamefont {V.}~\bibnamefont
  {Vovchenko}}, \bibinfo {author} {\bibfnamefont {A.}~\bibnamefont {Pasztor}},
  \bibinfo {author} {\bibfnamefont {Z.}~\bibnamefont {Fodor}}, \bibinfo
  {author} {\bibfnamefont {S.~D.}\ \bibnamefont {Katz}}, \ and\ \bibinfo
  {author} {\bibfnamefont {H.}~\bibnamefont {Stoecker}},\ }\href {\doibase
  10.1016/j.physletb.2017.10.042} {\bibfield  {journal} {\bibinfo  {journal}
  {Phys. Lett.}\ }\textbf {\bibinfo {volume} {B775}},\ \bibinfo {pages} {71}
  (\bibinfo {year} {2017})},\ \Eprint {http://arxiv.org/abs/1708.02852}
  {arXiv:1708.02852 [hep-ph]} \BibitemShut {NoStop}%
\bibitem [{\citenamefont {Vovchenko}\ \emph
  {et~al.}(2018{\natexlab{a}})\citenamefont {Vovchenko}, \citenamefont
  {Steinheimer}, \citenamefont {Philipsen},\ and\ \citenamefont
  {Stoecker}}]{Vovchenko:2017gkg}%
  \BibitemOpen
  \bibfield  {author} {\bibinfo {author} {\bibfnamefont {V.}~\bibnamefont
  {Vovchenko}}, \bibinfo {author} {\bibfnamefont {J.}~\bibnamefont
  {Steinheimer}}, \bibinfo {author} {\bibfnamefont {O.}~\bibnamefont
  {Philipsen}}, \ and\ \bibinfo {author} {\bibfnamefont {H.}~\bibnamefont
  {Stoecker}},\ }\href {\doibase 10.1103/PhysRevD.97.114030} {\bibfield
  {journal} {\bibinfo  {journal} {Phys. Rev.}\ }\textbf {\bibinfo {volume}
  {D97}},\ \bibinfo {pages} {114030} (\bibinfo {year} {2018}{\natexlab{a}})},\
  \Eprint {http://arxiv.org/abs/1711.01261} {arXiv:1711.01261 [hep-ph]}
  \BibitemShut {NoStop}%
\bibitem [{\citenamefont {Vovchenko}\ \emph
  {et~al.}(2018{\natexlab{b}})\citenamefont {Vovchenko}, \citenamefont
  {Steinheimer}, \citenamefont {Philipsen}, \citenamefont {Pasztor},
  \citenamefont {Fodor}, \citenamefont {Katz},\ and\ \citenamefont
  {Stoecker}}]{Vovchenko:2018zgt}%
  \BibitemOpen
  \bibfield  {author} {\bibinfo {author} {\bibfnamefont {V.}~\bibnamefont
  {Vovchenko}}, \bibinfo {author} {\bibfnamefont {J.}~\bibnamefont
  {Steinheimer}}, \bibinfo {author} {\bibfnamefont {O.}~\bibnamefont
  {Philipsen}}, \bibinfo {author} {\bibfnamefont {A.}~\bibnamefont {Pasztor}},
  \bibinfo {author} {\bibfnamefont {Z.}~\bibnamefont {Fodor}}, \bibinfo
  {author} {\bibfnamefont {S.~D.}\ \bibnamefont {Katz}}, \ and\ \bibinfo
  {author} {\bibfnamefont {H.}~\bibnamefont {Stoecker}},\ }in\ \href@noop {}
  {\emph {\bibinfo {booktitle} {{27th International Conference on
  Ultrarelativistic Nucleus-Nucleus Collisions (Quark Matter 2018) Venice,
  Italy, May 14-19, 2018}}}}\ (\bibinfo {year} {2018})\ \Eprint
  {http://arxiv.org/abs/1807.06472} {arXiv:1807.06472 [hep-lat]} \BibitemShut
  {NoStop}%
\bibitem [{\citenamefont {Almasi}\ \emph {et~al.}(2018)\citenamefont {Almasi},
  \citenamefont {Friman}, \citenamefont {Morita}, \citenamefont {Lo},\ and\
  \citenamefont {Redlich}}]{Almasi:2018lok}%
  \BibitemOpen
  \bibfield  {author} {\bibinfo {author} {\bibfnamefont {G.~A.}\ \bibnamefont
  {Almasi}}, \bibinfo {author} {\bibfnamefont {B.}~\bibnamefont {Friman}},
  \bibinfo {author} {\bibfnamefont {K.}~\bibnamefont {Morita}}, \bibinfo
  {author} {\bibfnamefont {P.~M.}\ \bibnamefont {Lo}}, \ and\ \bibinfo {author}
  {\bibfnamefont {K.}~\bibnamefont {Redlich}},\ }\href@noop {} {\  (\bibinfo
  {year} {2018})},\ \Eprint {http://arxiv.org/abs/1805.04441} {arXiv:1805.04441
  [hep-ph]} \BibitemShut {NoStop}%
\bibitem [{\citenamefont {Andronic}\ \emph {et~al.}(2018)\citenamefont
  {Andronic}, \citenamefont {Braun-Munzinger}, \citenamefont {Redlich},\ and\
  \citenamefont {Stachel}}]{Andronic:2017pug}%
  \BibitemOpen
  \bibfield  {author} {\bibinfo {author} {\bibfnamefont {A.}~\bibnamefont
  {Andronic}}, \bibinfo {author} {\bibfnamefont {P.}~\bibnamefont
  {Braun-Munzinger}}, \bibinfo {author} {\bibfnamefont {K.}~\bibnamefont
  {Redlich}}, \ and\ \bibinfo {author} {\bibfnamefont {J.}~\bibnamefont
  {Stachel}},\ }\href {\doibase 10.1038/s41586-018-0491-6} {\bibfield
  {journal} {\bibinfo  {journal} {Nature}\ }\textbf {\bibinfo {volume} {561}},\
  \bibinfo {pages} {321} (\bibinfo {year} {2018})},\ \Eprint
  {http://arxiv.org/abs/1710.09425} {arXiv:1710.09425 [nucl-th]} \BibitemShut
  {NoStop}%
\bibitem [{\citenamefont {Andronic}(2014)}]{Andronic:2014zha}%
  \BibitemOpen
  \bibfield  {author} {\bibinfo {author} {\bibfnamefont {A.}~\bibnamefont
  {Andronic}},\ }\bibfield  {booktitle} {\emph {\bibinfo {booktitle}
  {{Proceedings, 26th International Symposium on Lepton Photon Interactions at
  High Energy (LP13)}}},\ }\href {\doibase 10.1142/S0217751X14300476}
  {\bibfield  {journal} {\bibinfo  {journal} {Int. J. Mod. Phys.}\ }\textbf
  {\bibinfo {volume} {A29}},\ \bibinfo {pages} {1430047} (\bibinfo {year}
  {2014})},\ \Eprint {http://arxiv.org/abs/1407.5003} {arXiv:1407.5003
  [nucl-ex]} \BibitemShut {NoStop}%
\bibitem [{\citenamefont {Bzdak}\ and\ \citenamefont
  {Koch}(2018)}]{Bzdak:2018axe}%
  \BibitemOpen
  \bibfield  {author} {\bibinfo {author} {\bibfnamefont {A.}~\bibnamefont
  {Bzdak}}\ and\ \bibinfo {author} {\bibfnamefont {V.}~\bibnamefont {Koch}},\
  }\href@noop {} {\  (\bibinfo {year} {2018})},\ \Eprint
  {http://arxiv.org/abs/1811.04456} {arXiv:1811.04456 [nucl-th]} \BibitemShut
  {NoStop}%
\bibitem [{\citenamefont {Luo}(2015{\natexlab{b}})}]{Luo:2015ewa}%
  \BibitemOpen
  \bibfield  {author} {\bibinfo {author} {\bibfnamefont {X.}~\bibnamefont
  {Luo}} (\bibinfo {collaboration} {STAR}),\ }\bibfield  {booktitle} {\emph
  {\bibinfo {booktitle} {{Proceedings, 9th International Workshop on Critical
  Point and Onset of Deconfinement (CPOD 2014): Bielefeld, Germany, November
  17-21, 2014}}},\ }\href@noop {} {\bibfield  {journal} {\bibinfo  {journal}
  {PoS}\ }\textbf {\bibinfo {volume} {CPOD2014}},\ \bibinfo {pages} {019}
  (\bibinfo {year} {2015}{\natexlab{b}})},\ \Eprint
  {http://arxiv.org/abs/1503.02558} {arXiv:1503.02558 [nucl-ex]} \BibitemShut
  {NoStop}%
\end{thebibliography}%

\end{document}